\documentclass[journal]{IEEEtran}

\usepackage{times}
\usepackage{multirow}

\ifCLASSINFOpdf
   \usepackage[pdftex]{graphicx}
\else
\usepackage[dvips]{graphicx}

 \usepackage{amssymb,amsmath}
\fi

\usepackage{subfig}
\usepackage{cite}
\usepackage{caption}
\usepackage{url}
\usepackage{array}
\usepackage{colortbl}
\usepackage{tabulary}
\usepackage{etoolbox}
\usepackage[dvipsnames]{xcolor}

\begin{document}

\title{BLE Beacons for Indoor Positioning at an Interactive IoT-Based Smart Museum}

\author{Petros Spachos,~\IEEEmembership{Senior Member,~IEEE,}
        and~Konstantinos~N.~Plataniotis,~\IEEEmembership{Fellow,~IEEE}
        
\thanks{This work was supported in part by the Natural Sciences and Engineering Research Council of Canada (NSERC). }        
\thanks{ P. Spachos is with the School of Engineering, University of Guelph, Guelph, ON, Canada. (e-mail:  \textit{petros}@uoguelph.ca).

Konstantinos N. Plataniotis is with the Department of Electrical and Computer Engineering, University of Toronto, Toronto, ON, Canada. (kostas@ece.utoronto.ca)}% <-this % stops a space

        }

\maketitle

\begin{abstract}
The Internet of Things (IoT) can enable smart infrastructures to provide advanced services to the users. New technological advancement can improve our everyday life, even simple tasks as a visit to the museum. In this paper, an indoor localization system is presented,  to enhance the user experience in a museum.  In particular, the proposed system relies on  Bluetooth Low Energy (BLE) beacons proximity and localization capabilities to automatically provide the users with cultural contents
related to the observed artworks. At the same time, an RSS-based technique is used to estimate the location of the visitor in the museum. An Android application is developed to estimate the distance from the exhibits and collect useful analytics regarding each visit and provide a recommendation to the users. Moreover, the application implements a simple Kalman filter in the smartphone, without the need of the Cloud, to improve localization precision and accuracy. Experimental results on distance  estimation, location, and detection accuracy show that BLE beacon is a promising solution for an interactive smart museum. The proposed system has been designed to be easily extensible to the IoT technologies and its effectiveness has been evaluated through experimentation.\end{abstract}


\IEEEpeerreviewmaketitle

\section{Introduction}
\IEEEPARstart{O}{ver} the centuries, the traditional role of museums is to collect objects and materials of cultural, religious and historical importance, preserve them, research into them and present them to the public for the purpose of education and enjoyment. Nowadays, museums usually provide the visitor with paper booklets or with audio guides to help them navigate in the large museum areas.  These booklets are designed for the general audience and sometimes they fail to meet the needed of individuals with special interest. At the same time, as the number of exhibits and collections increases, while the available time of the visitors is limited, they might not manage to visit the exhibits they are interested in or even explore more exhibits and collections related to their personal interest. As a result, sometimes visits to the museum ends up being  boring or even too difficult  for many visitors. Therefore, an interactive and personalized museum tour that takes into consideration the available time for a visit and the personal interests of the visitors is needed.

A popular way to contain exhibits and collection information is the use of Quick Response (QR) code~\cite{schultz}. Visitors can get a brief introduction, images or even a website with related information by scanning the QR code on their mobile device. Another approach includes Augmented Reality (AR), where the visitors can interact with objects~\cite{bimbo, tsai}. However, both approaches require the visitors to take action in order to get further information about an exhibit, either by scanning the QR code or by downloading all the necessary software and application for the AR technology. Passive approaches can be helpful for visitors that are already interested in an exhibit and know where to find it, but  they are not helpful  for the majority of the visitors that have limited or no knowledge at all about the different available collections. 

Positioning technologies, especially those based on the proximity from the different objects can help and increase the  interaction with the visitors in an active way, where the exhibit triggers an action. Location-aware services can guide the visitor in a complex indoor environment such as a large museum. However, the deployment of an accurate indoor location-aware system is challenging, while it might not be available in every object of interest. At the same time, museums tend to change the location of the exhibits and collections over time, making it difficult to have some basic coordination for the localization system. A system that does not use the absolute location but relative information regarding the object within a range from the visitor can be more useful.

The increased popularity of smartphones along with the development of the Internet of Things (IoT) can alleviate the problem. Low-cost and small size devices can interact with a smartphone application and provide useful information to the visitor, without affecting the current infrastructures at the museum. IoT devices can work autonomously with minimal human intervention to provide a simple service. Bluetooth Low Energy (BLE) beacons, commonly referred to as beacons, are small wireless devices that can provide proximity services to nearby BLE enabled devices~\cite{jeon, spachosmicro}. Beacons can be placed almost in any indoor location and convert a traditional room into a smart environment where the 
visitor interacts with the objects based on her/ his distance from them.
 
In this paper, the proposed system uses  BLE beacons for an IoT-based smart museum. The visitors use a developed Android application that runs in the background. When the visitors are close to an exhibit, they receive a notification about the exhibit. Then, the user decides either to get more information about the exhibit or ignore it. While the visitor uses the application, her/ his location is estimated based on the  Received Signal Strength Indicator (RSSI) values of nearby BLE beacons. At the same time, the proposed system captures useful analytics about the visitors' retention time - the time they spend in the museum,  and interest towards the different collections and  exhibits, and based on this information to provide a recommendation for future visits.

To evaluate the performance of the proposed system three sets of experiments were conducted in two environments: In the first experiment, the distance estimation accuracy of a single BLE beacon was examined. In
the second experiment, the localization performance of three BLE beacons was examined, when the receiver is moving between them. In the third experiment, the  detection accuracy of the developed application is examined when three beacons are placed side by side in different topologies. The experimental results verified some expected assumption, but also revealed some interesting insights. According to the experimental results:

\begin{itemize} 
\item Beacons can be placed anywhere without interfering with any other wireless infrastructures in the area. When needed, the location of each beacon can also change easily. However, an accurate path loss model of the  deployment  area is necessary for the beacons to have acceptable performance. 

\item The location accuracy of the beacons in a complex indoor environment is sufficient for an application such as the smart museum when errors within a few meters  might be acceptable. Advanced filtering techniques can improve the accuracy. Location accuracy increases as the distance between the beacons that act as anchors increases.

\item When the receiver is close to the beacon, the detection estimation accuracy is acceptable. As the receiver is moving further away, the estimation accuracy decreases. At the same time, as the distance between neighboring beacons increases the detection estimation accuracy is more challenging. Also, as the number of the neighboring beacons increases, the detection estimation accuracy decreases.

\item Any BLE-based application should take into consideration the unique characteristics of the deployment area, such as noise and interference. Beacon's performance is affected by many factors, however, their proper placement can improve the system accuracy.
\end{itemize}

The rest of this paper is organized as follows: Section~\ref{related} reviews the related works, followed by an indoor localization technique based on iBeacons in Section~\ref{indloc}. The proposed system architecture is discussed in Section~\ref{arch} followed by a number of experiments to evaluate the performance of the system in Section~\ref{exp}. The conclusions are in Section~\ref{concl}.

\section{Related Work} \label{related}
Wireless indoor positioning systems have become very popular in recent years~\cite{liu, sadowski}. One promising technology is Ultrawide Bandwidth (UWB). UWB can provide accurate localization capabilities through Time-Of-Arrival (TOA)-based ranging techniques~\cite{dardari}. UWB is  a great candidate for indoor localization due to its power efficiency, fine delay resolution, and robust operation in harsh environments~\cite{marano}. However, UWB requires extra hardware devices to be deployed.  
 
RSS- based indoor positioning systems are more popular due to their availability and  low cost~\cite{sadowski, feng, mazuelas, pivato, tomic}. Although RSSI is prone to noise and interference, there are techniques that can improve its performance when it comes to localization. In~\cite{mackey1}, Bayesian filters are used to improve estimation accuracy. In~\cite{paul}, a sigma-point Kalman smoother (SPKS)-based location and tracking algorithm is proposed. SPKS algorithm has higher accuracy in comparison with a commercially available positioning engine, over a number of trials. In~\cite{chen}, an improved unscented Kalman filter and the particle swarm optimization (PSO) are proposed. PSO can reduce the positioning error and improve positioning accuracy. In~\cite{xue}, the average of a number of selected maximum RSSI observations is used to improve the accuracy. Experiments were conducted in four rooms and a corridor within an office building  with promising results in positioning accuracy. In~\cite{qwang},  a Gaussian Mixture Model (GMM) to model the distribution of a set of Non-Line-Of-Sight (NLOS) corrupted range estimations is proposed. In~\cite{stomic}, convex optimization is used to address the RSS-based noncooperative and cooperative localization problems, while a Linear Least Squares (LLS) estimator is proposed in~\cite{cso}. In~\cite{zhang}, a multi-task correlation particle filter for robust visual tracking is proposed, while in~\cite{imani} particle filters for partially-observed Boolean dynamical systems are examined.  A multi-fidelity Bayesian optimization algorithm for the inference of general nonlinear state-space models is proposed in~\cite{imani3} and a Bayesian decision framework in~\cite{imani2}.
 
With the increased popularity of the IoT,  people in their daily lives are surrounded by more and smarter devices such as laptops, smartphones, and tablets that are capable of collecting RSSI signals~\cite{tarkoma, stankovic}. BLE beacons are small size, low-cost devices that can be used  for indoor localization~\cite{huh, molina}.  IoT devices can be found in a plethora of application for smart cities~\cite{jjin, zanella}, smart homes~\cite{kelly}, healthcare~\cite{islam} and many more~\cite{al-fuqaha,dxu,he}. At the same time, the popularity of smartphone and mobile devices have also enabled the smartphone-based indoor localization where a number of sensors are used to measure human mobility and enrich location context~\cite{yang}. 

When it comes to smart museums, there are a few recent approaches that try to increase interaction and provide analytics regarding visitors' retention time~\cite{yoshimura,alletto, jimenez}. In~\cite{vassilakis}, a framework that allows for self-aware exhibits positioned close to each other to cooperate and work together, to produce self-organized exhibitions is proposed. In~\cite{haworth}, a smartphone is used  to follow trails in a museum by scanning QR codes.  In~\cite{karimi},  RFID-Enhanced Museum for Interactive Experience (REMIX) which aims to develop a personalization platform for museums based on RFID technology  is presented. An automatic museum guide system that provides both interactive guidance for exhibition and NFC-based location navigation is presented in~\cite{cai}. A noninvasive Bluetooth monitoring of visitors' length is proposed in~\cite{yoshimura}, using the visitor's mobile devices to get useful analytics.  In~\cite{alletto}, a wearable device that combines image recognition and localization capabilities through BLE beacons is proposed. The system,  apart from the localization algorithm, it also requires images to provide accurate information to the visitor. In~\cite{jimenez}, a location-aware service for a museum is proposed. The authors examine both UWB and BLE techniques  and combine them with Pedestrian Dead-reckoning (PDR) estimation and they demonstrate that BLE ranging techniques along with smartphone-based PDR is feasible in a museum-like use case.  

In this work, BLE beacons are used along with a developed Android application. Beacons are the only source of information to extract the location of the visitor and any other analytics regarding their visit. No other sensor or devices are used, making the deployment of the proposed system easier while increasing its usability and applicability.
  
\section{Indoor Localization Based on Beacons}\label{indloc}
In this section, the main features of the BLE beacons are described, followed by the ranging technique that is used for the proximity estimation. Then, trilateration which is used for localization is briefly described.

\subsection{BLE beacons features}
BLE beacons, usually referred to as beacons, are small, inexpensive, battery-operated  wireless transmitters~\cite{jeon, spachosmicro}, and they can have several protocols. In this work, the iBeacon protocol is used~\cite{ibeaconpacket}. Beacons broadcast their identifier to nearby electronic devices that support BLE signals, such as smartphones or single-board computers.   They use only advertising mode, which is one-way BLE discovery process. They periodically send packets of data that can be received by other devices like smartphones or tablets.  They only send a signal and not listening. Their signal can be transmitted at intervals from 20~ms  up to 10~s. The transmission interval affects the battery life  of the beacons. The longer the interval, the more available battery left.

\subsubsection{BLE wireless technology}
BLE beacons use  BLE technology. BLE was introduced by the Bluetooth Special Interest Group as a subsystem of Bluetooth, in order to achieve device discovery and connectivity with low power consumption~\cite{blue4}. BLE was designed for applications that do not need to exchange a large amount of data and intend to provide considerably  reduced power consumption and cost, while it maintains similar communication range with classic Bluetooth. BLE is popular among IoT devices due to its low-cost and low-power requirements.  
 
BLE 4.0 can  reach 25~Mbit/s at a distance of 60~m. Although BLE and Wi-Fi utilize the same radio frequency bands, BLE advertising only occurs on three channels, 37, 38 and 39,  and they are widely spaced at 2402~MHz, 2426~MHz, and 2480~MHz, which  separates them from the popular Wi-Fi channels. In this way, BLE prevents interference with other Wi-Fi infrastructures in the deployment area.  Unfortunately, only the Received Signal Strength (RSS) values are registered but not the channel on which the packet was received, leading to important fading of 30~dB in very close positions~\cite{Faragher}. 

Simplicity and popularity of BLE among IoT devices made it one of the promising technologies or microlocation~\cite{spachosmicro}.  There are a plethora of energy-efficient, low-cost BLE beacon vendors that build beacons to meet the needs of various applications. Among the disadvantages is that BLE is prone to interference, however, there are techniques that can be used to minimize it.
  
\subsubsection{BLE beacons characteristics}

Beacons have several characteristics that made them a promising solution for indoor localization technology. Their small size made it possible to place them almost anywhere in a complex environment, without disturbing other infrastructures. They can work for months with single coin cell batteries~\cite{gimbal}, while there are also beacons with two AA batteries~\cite{bluecats} for an extended lifetime or even USB-powered beacons and solar-powered beacons~\cite{cyalkit, spachos1} for outdoor deployment.  Beacons can be located behind or on the side of an item and send notifications about it to other devices in the range.

Another important characteristic of beacons that works as a tradeoff is their transmission power. As every wireless transmitting device, the transmission power directly affects the transmission range. Beacons can reach up to 60~m of transmission however, this will drain their battery faster, while it can create interference to other beacon transmissions in the area. A transmission range between 2 to 5~m is enough for an application as the proposed smart museum.
  
The time between consecutive transmissions, known as advertising interval, is also important for the lifespan of the beacons. When  the receivers are moving fast in the area, a short advertising interval is necessary but unfortunately, the signal might not be stable. A longer advertising interval will provide a more stable signal and extend beacon's battery lifetime, while it might not reach fast-moving receivers. Advertising interval is another tradeoff when designing an application using beacons.
  
The main characteristic of beacons that makes them ideal for indoor localization is their measured power. This is the expected RSS at 1~m distance  from the beacon. The receivers can use this value, calibrate it and eventually find the distance from the beacon. There are many works in the literature that focus on RSS-based localization~\cite{sadowski, paul, Zanca, luo}.

\subsection{Ranging technology}

The RSSI from the beacon can be used to find the distance between the beacon and the receiver. RSS-based localization is among the most popular techniques for localization due to its simplicity and signal availability. As the radio wave propagates from the beacon to the receiver according to the inverse-square law, the distance between the two can be calculated, as long as no other errors contribute to faulty results.

Each beacon sends its location ID along with the Transmission Power (TX) value. In its simplest form, the path loss can be calculated using the formula:

\begin{equation}
RSSI =-10n\log_{10}d + A
\label{rssimain}
\end{equation}
where $n$ is a signal propagation constant depending mainly on the environment, $d$ is the distance,  and $A$ is the received signal strength at 1~m.

Following the traditional path loss model, the formula for a noisy environment ~\cite{mazuelas}:

\begin{equation}
RSSI = RSSI_{0} - 10n\log_{10} \left( \frac{d}{d_0}\right) + v
\end{equation}
where $d_0$ is the reference distance in 1~m, $RSSI_0$ is the mean $RSSI$ value obtained at the reference distance of $d_0$, $n$ is the path loss component and $v$ is a Gaussian random variable with zero mean and standard deviation $\sigma_{RSSI}$ that accounts for the random effect of shadowing. 

When solving for the distance and without taking into account RSSI noise:

\begin{equation}
d_{noiseless}=d_0 10^{\frac{RSSI_0-RSSI}{10n}}
\end{equation}

However, for a real noisy RSSI with deviation $\sigma_{RSSI}$ the estimated distance is given by~\cite{Zanca}:

\begin{equation}
d=d_{noiseless} \exp^{-0.5\left( \frac{\sigma_{RSSI}\ln10}{10n} \right)^2}
\end{equation}

As it can be seen, the accuracy of the calculation of the path loss component can affect the accuracy of the system. During experimentation, knowing $RSSI_0$ in $d_0$ can help to calculate $n$ from Eq.~(\ref{rssimain}).

\subsection{Trilateration}
A popular indoor positioning technique is lateration. Lateration is the process of estimating the location of the receiver, given the distance from a set of points with know location.

\begin{figure}[!t]
\centering%
\includegraphics[width=0.8\columnwidth]{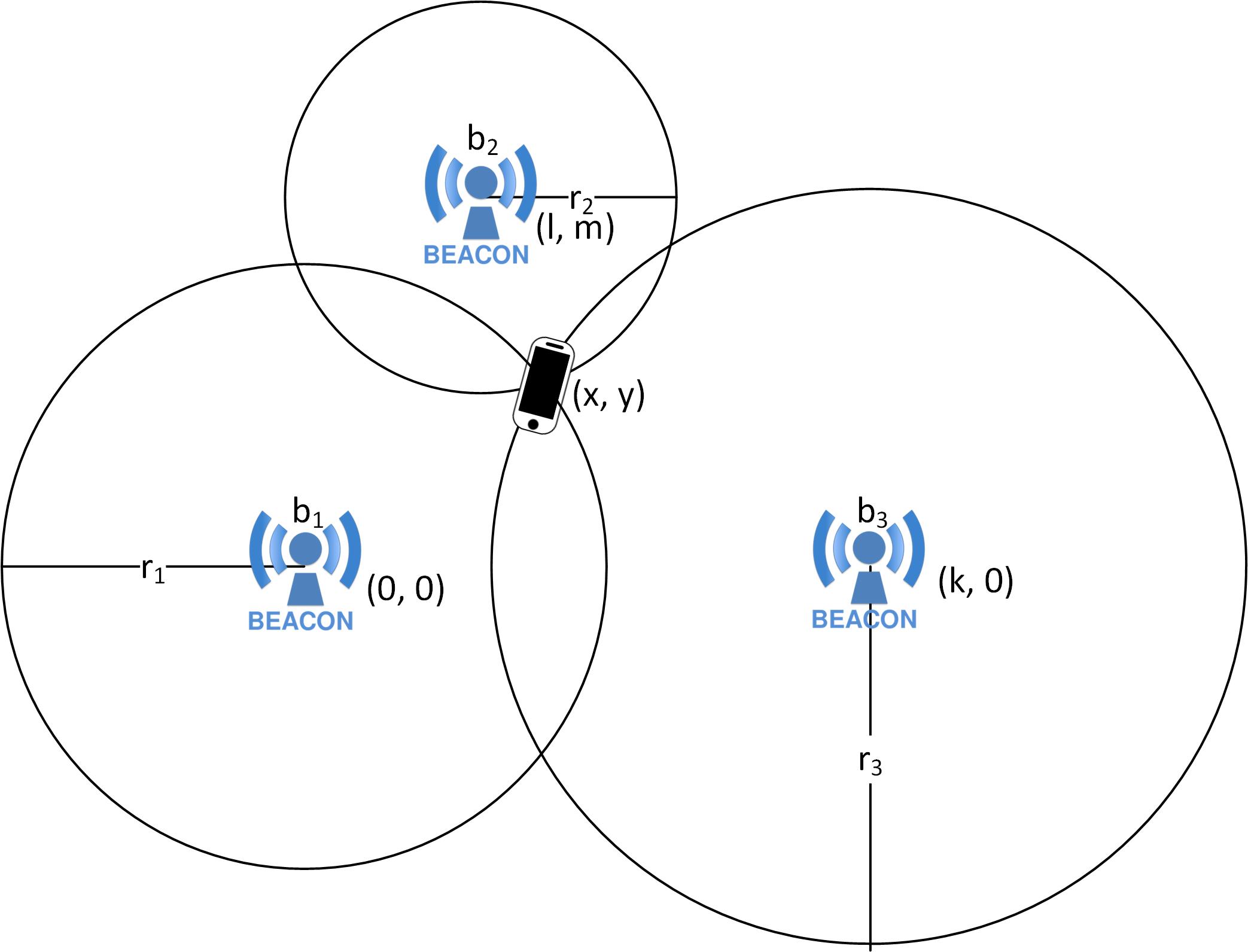}%
\caption{An example of trilateration with three beacons, $b_1, b_2$ and $b_3$ in known locations, (0, 0), (l,m) and (k,0), respectively, are the transmitters and a smartphone at the intersection, (x, y), as the receiver.}%
\label{trilateration}%
\end{figure}

An example of lateration with three known points, trilateration, is shown in Fig.~\ref{trilateration}. Trilateration calculates the intersecting point of the three circles, where the smartphone is located when the center points of each beacon and their radii are already known. The radii can be determined from Eq.~(\ref{rssimain}). 

Assuming that the smartphone is  located at $(x, y)$, beacon 1 at $b_1=(0, 0)$, beacon 2 at $b_2=(l, m)$, and beacon 3 at $b_3=(k, 0)$ and the radii  $r_1, r_2$  are  known, then:

\begin{equation}
r_1^2 = x^2 + y^2
\label{r1}
\end{equation}
\begin{equation}
r_2^2 =  (x-l)^2 + (y-m)^2
\label{r2}
\end{equation}
\begin{equation}
r_3^2 =  (x-k)^2 + y^2
\label{r3}
\end{equation}

Combining Eq.~(\ref{r1}), Eq.~(\ref{r2}), and Eq.~(\ref{r3}), the location of the smartphone can be calculated as:

\begin{equation}
x= \frac{r_1^2 - r_3^2 +k^2}{2k}
\end{equation}
\begin{equation}
y= \frac{r_1^2 - r_2^2 + l^2 +m^2}{2m} - \frac{l}{m}x
\end{equation}

Since RSSI is prone to interference and noise, it is expected an error in the calculation of the actual location. The error between the estimated and the real location can be found using the Mean Square Error (MSE):

\begin{equation}
MSE_{est}= \sqrt{(x_{est} - x_{real})^2 + (y_{est} - y_{real})^2 }
\end{equation}
 
MSE helps calculate the accuracy of the system and perform any necessary calibrations.

\section{Proposed System Architecture}\label{arch}
The proposed IoT-based smart museum has three main services. Along with a number of beacons that are located in different spots in the museum, an Android application that is installed in a visitor's smartphone and a data collection platform running on the server of the museum.

\subsection{System services}

An illustration of the system services is shown in Fig.~\ref{system}.
The main services offered by the proposed systems are the following:

\begin{figure}[!t]
\centering%
\includegraphics[width=\columnwidth]{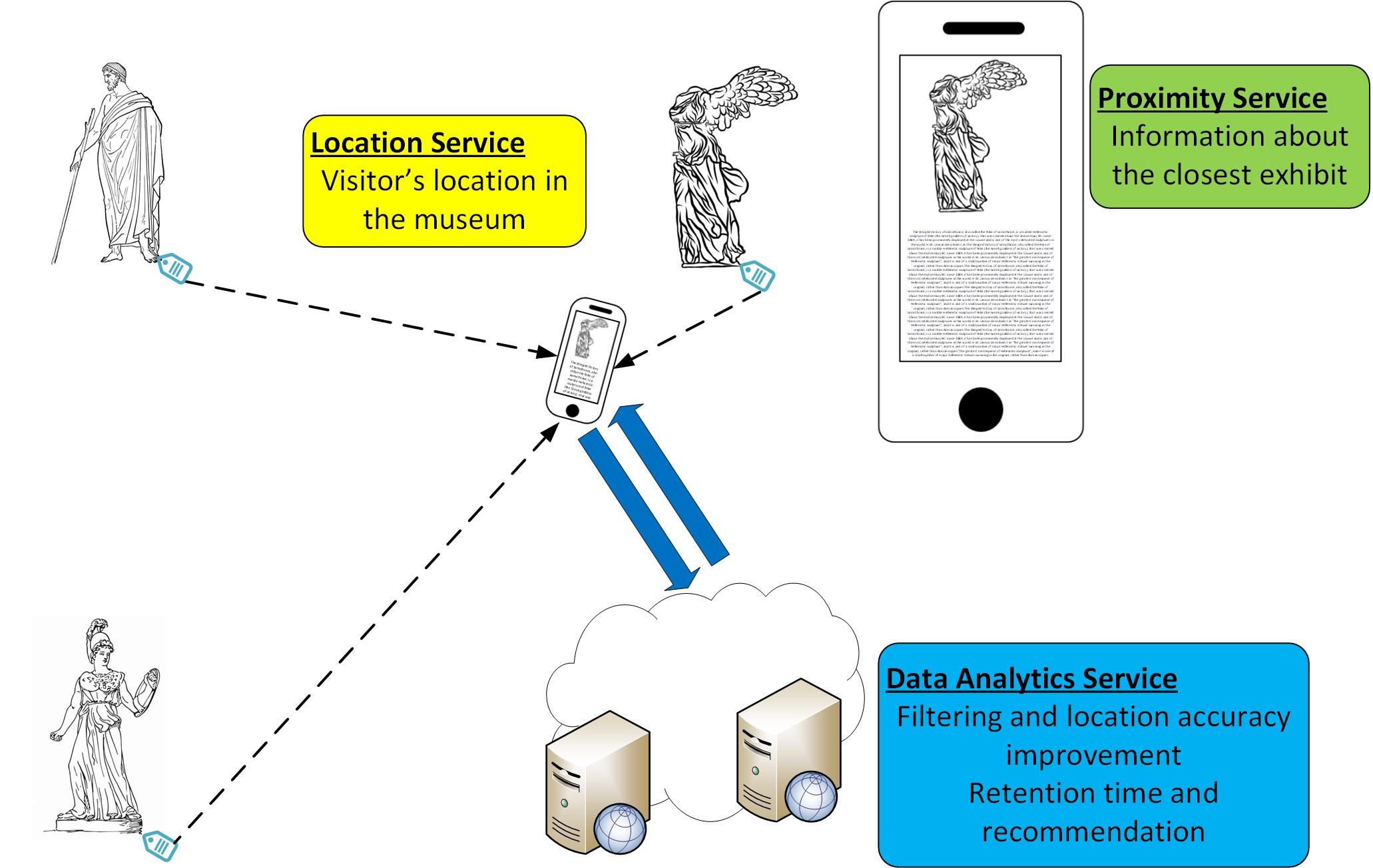}%
\caption{Illustration of the system services. There are three main services: The location service, the proximity service, and the data analytics service.}%
\label{system}%
\end{figure}

\begin{enumerate}
\item \textit{Proximity service.} Beacons at each exhibit broadcast their messages continuously. When the visitor is close to an exhibit, she/ he gets a notification about the exhibit on the smartphone. If the visitor gets closer to the exhibit, more information regarding the specific exhibit will be forward to the smartphone. At the same time, the interest of the visitor at the specific exhibit is recorded at the application.

\item \textit{Localization service.} As the visitor is moving between the beacons, the different messages coming  from the beacons are used to provide an approximate location of the visitor in the museum. This information can be used from the visitor to navigate to the different rooms of the museum and find the exhibits she/ he is interested in. The application running on the smartphone can also track the path of the visitor in the museum and provide a recommendation based on the visitor's preferences and distance from the different rooms.

\item \textit{Data analytics service.}  The path that the visitor followed along with her/ his  preferences, the retention time and the timestamp of each action are forwarded to a processing center. These data are collected and provide useful analytics for the visitor and the management of the museum. The visitor can mark the exhibits she/ he visited along with any notes she/ he made. At the same time, in the application, the visitor can set up a path with the desired exhibits. The management of the museum collects analytics about the  number of visitors in each exhibit and their retention time throughout the day. These can be useful information to improve the visibility of some exhibits, make sure to be able to accommodate a large number of visitors in specific exhibits during rush hours and provide  a recommendation to the visitors according to their interest.

\end{enumerate}

If the museum offers Wi-Fi access  all the above services can benefit from real-time access to a devoted server for the application. The location estimation can be improved through filtering such as Kalman or Particle filters, while the recommendation for suggested iteration can be received from the visitor in real-time, based on the retention time at each exhibit.

\subsection{Hardware components}
\begin{figure}[t!]
    \centering
    \includegraphics[width=0.6\columnwidth]{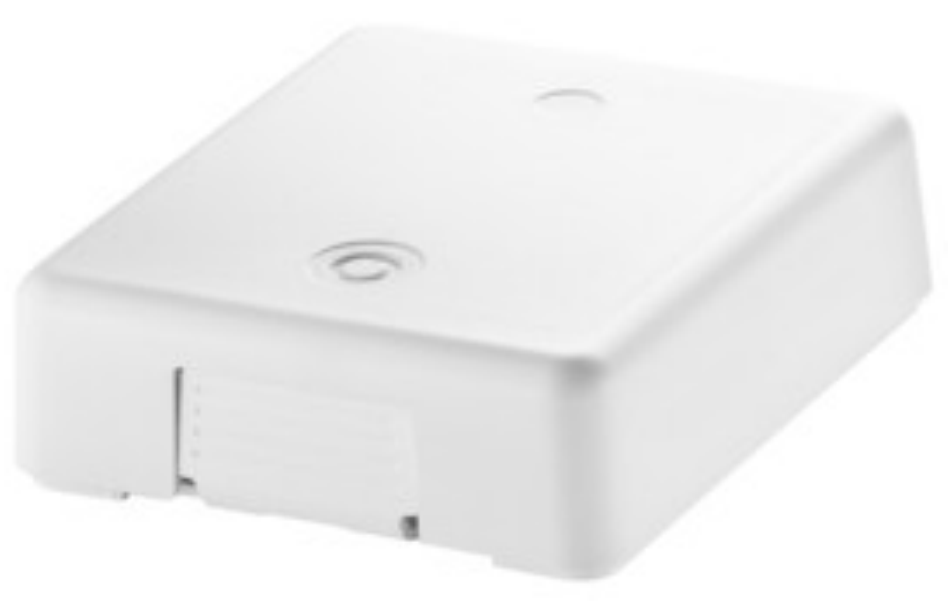}%
    \captionof{figure}{Gimbal Series 21.}
    \label{gimbal}
  \end{figure}

There are a plethora of beacon vendors in the market. There are beacons in different sizes, with different power sources and a different number of sensors.  In the proposed system, the Gimbal Series 21 beacons were used~\cite{gimbal} due to their low price and extended lifetime. Series 21 beacons they use four AA  batteries and have a typical battery life of 18 months, transmitting every 100~ms running 24~hours/ day~\cite{gimbal}. A Series~21 beacon  is shown in Fig.~\ref{gimbal} along with its specifications in Table~\ref{gimbalspec}.

As a receiver during experimentation, LG Nexus 5 was used. It has  Bluetooth 4.0, which is required for the developed application. During experimentation, the OS was Android~6.0.1.

\subsection{Software components}
An Android application was developed for the proposed system. The visitor needs to download  and install the application at the beginning of her/ his visit to the museum. The application requests for Bluetooth access and then it runs in the background.

When the visitor is close to a beacon, the application sends a notification to the visitor. If the visitor goes closer to the exhibit, the application comes in the foreground and start displaying information about the exhibit. While the visitor is in the proximity of the exhibit, the application records the retention time and the beacon ID. If there is a Wi-Fi connection, these data are forwarded to the control room, where the recommendation system is running. If there is no Wi-Fi connection, the application stores all the information locally and forwards them to the server the next time there is a wireless connection.  

   \begin{table} 
    \centering
        \normalsize

     \begin{tabular}[b]{|l|r|}\hline
             
             \textbf{Transmission Type} & Bluetooth 4.0 Low Energy \\ \hline
            \textbf{Antenna} & Omni-directional \\ \hline 

            \textbf{Transmission} & Configurable \\ 
            \textbf{Interval} &  from 100 ms up to 10 s \\ \hline
            \textbf{Transmission}  & Configurable  \\ 
            \textbf{Power} &  from -23 dBm up to 0 dBm \\ \hline
            \textbf{Transmission Range} & Typical up to 50 meters \\ \hline
                        \textbf{Battery Type} & 4 - AA Alkaline   \\ \hline
                        \textbf{Battery Life} & Typical up to 18 months\\ \hline  
                         \textbf{Dimensions} & 3.4 x 3.0 in x 1.0 in \\ 
            (L x W x H)  & (86 mm x 77 mm x 25 mm)\\ \hline 
            \multirow{2}{*}{\textbf{Weight}} & 170 grams  \\ 
            & (including batteries) \\ \hline

        \end{tabular}

      \captionof{table}{Beacon specifications.}
            \label{gimbalspec}
            \end{table}

\begin{figure}[t!]  
\centering
\captionsetup[subfloat]{farskip=0pt}%
\subfloat[Laboratory experiment.]{\includegraphics[width=0.8\columnwidth]{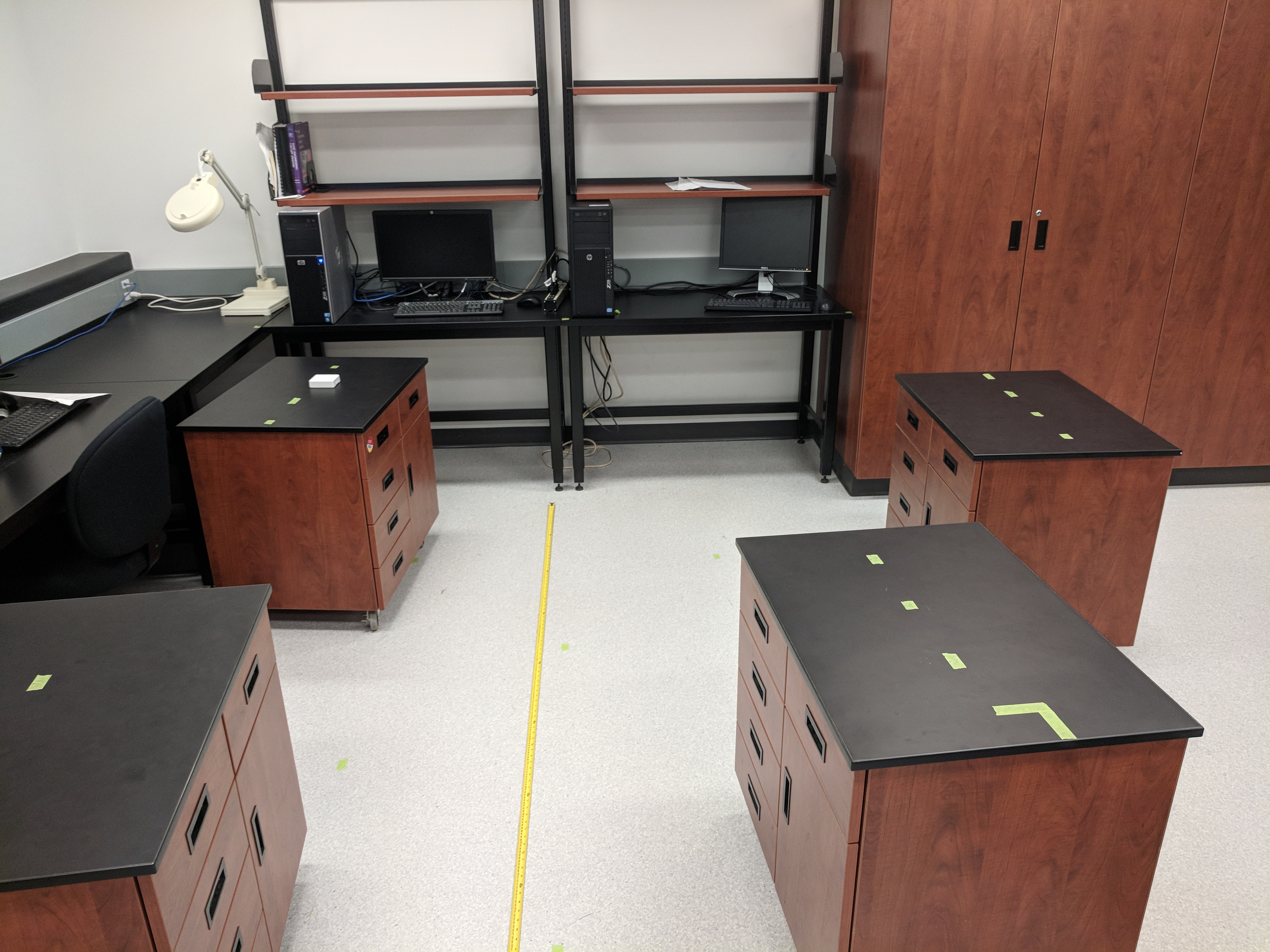}
\label{lab}}\vspace{2ex}
\subfloat[Corridor experiment.]{\includegraphics[width=0.8\columnwidth]{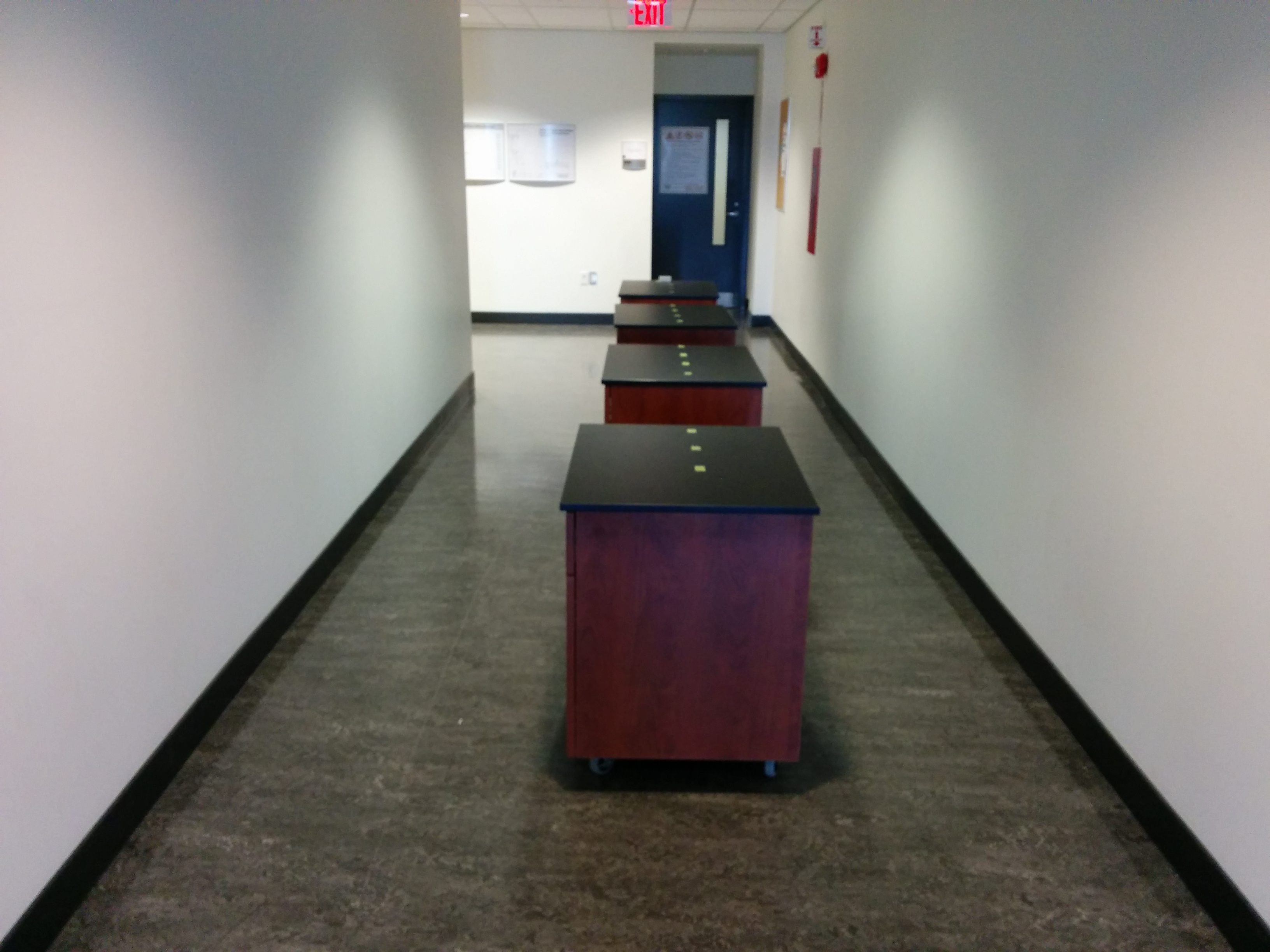}
\label{cor}}
\caption{Experimental environment.}
\label{labraw}
\end{figure}

The localization also runs on the application. Based on the RSSI from the neighboring beacons and their ID, the visitor's location on the museum's map is displayed. If a Wi-Fi is available, the accuracy of the localization can be improved with the use of particle filtering~\cite{mackey1}, while with no  Wi-Fi connection, Kalman filtering is running on the application~\cite{mackey}.

\section{Experimental Evaluation} \label{exp}
The performance of the proposed system was evaluated through experimentation in a large laboratory room, shown in Fig.~\ref{lab}, and at a corridor, shown in Fig.~\ref{cor}, at  our University. Three sets of experiments were conducted in each environment to evaluate the system under different environmental parameters, such as parallel wireless transmissions, number of people  in the area and  number and type of obstacles in the environment.

\subsection{Path loss model}

\begin{figure}[!t]
\centering
\captionsetup[subfloat]{farskip=0pt}%
\subfloat[Laboratory.]{\includegraphics[width=0.68\columnwidth]{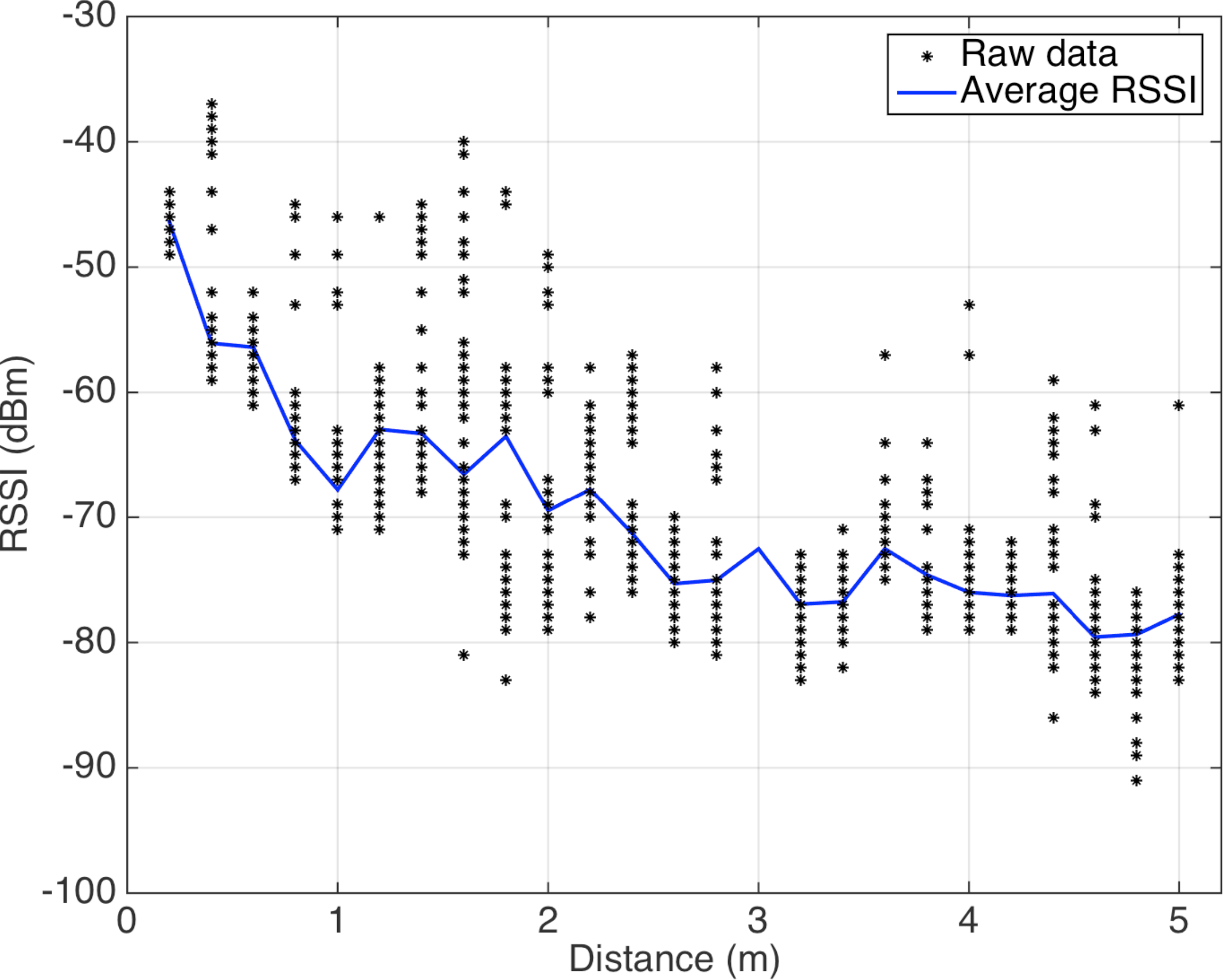}
\label{rawdata1}}\vspace{1ex}
\subfloat[Corridor.]{\includegraphics[width=0.68\columnwidth]{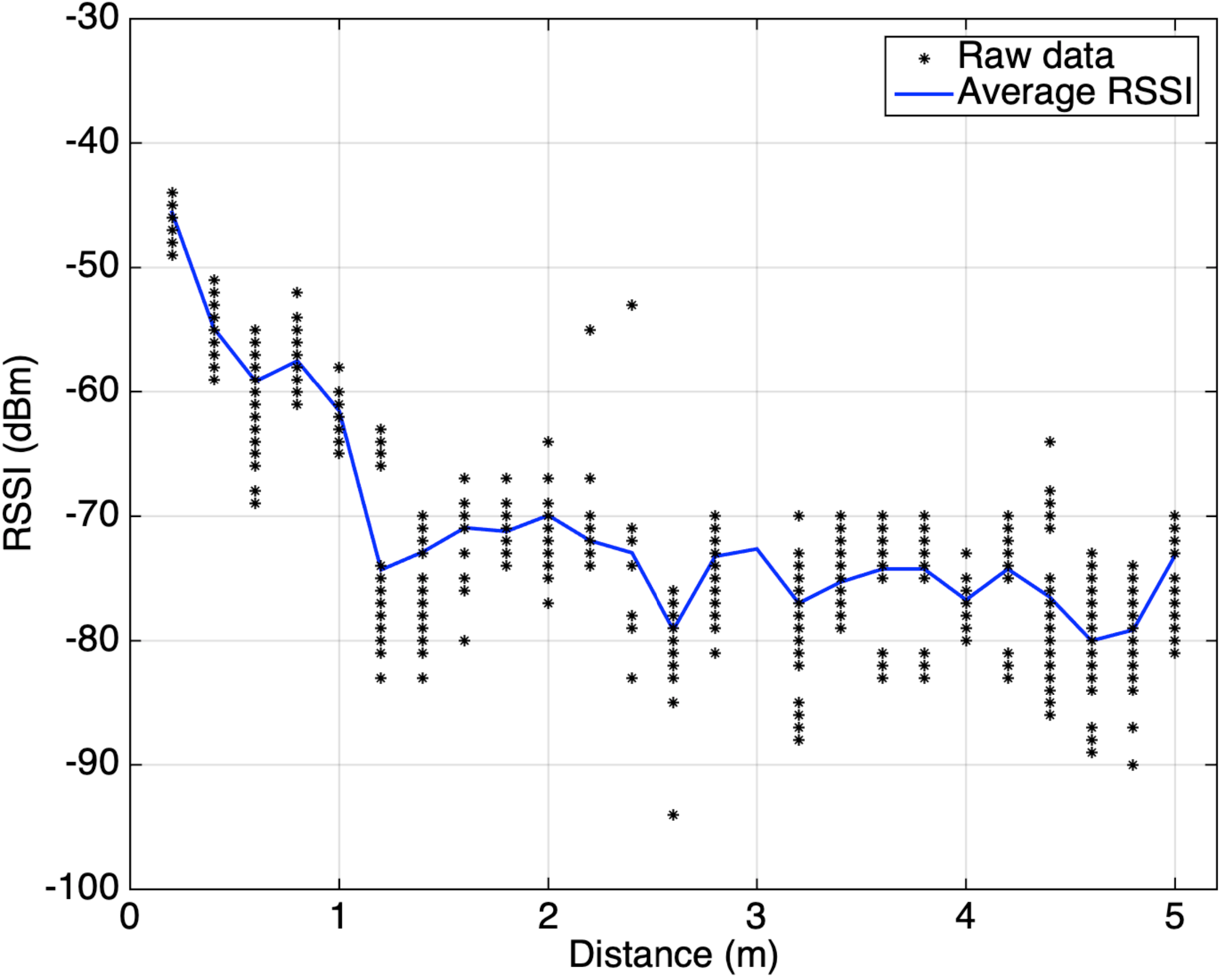}
\label{rawdata2}}
\caption{Raw RSSI values and average.}
\label{labraw}  
\end{figure}

At the beginning of the experimentation, the path loss model in each of the experimental environment was determined. A smartphone was used to collect the RSSI values every 20~cm and up to 5~m from the beacon. In each location, values were collected for 10~min.  The results for the laboratory and the corridor are shown in Fig.~\ref{rawdata1} and Fig.~\ref{rawdata2}, respectively.

In both environments, the RSSI value varies over time. The variation is greater as the distance between the beacon and the smartphone increases. Also, the RSSI variation in the corridor is higher than the RSSI variation in the laboratory. In the laboratory, the beacon transmits in a large area, away from walls or windows that can affect the signal. People are walking around but most of the time, there is a Line-of-sight (LOS) between the beacon and the smartphone. On the other hand, the corridor has concrete walls close to the beacon, hence space is limited. People walking on the corridor also block the signal between the beacon and the smartphone for most of the time, due to the limited space.

\begin{figure}[t!]  
\centering
\captionsetup[subfloat]{farskip=0pt}%
\subfloat[Laboratory.]{\includegraphics[width=0.68\columnwidth]{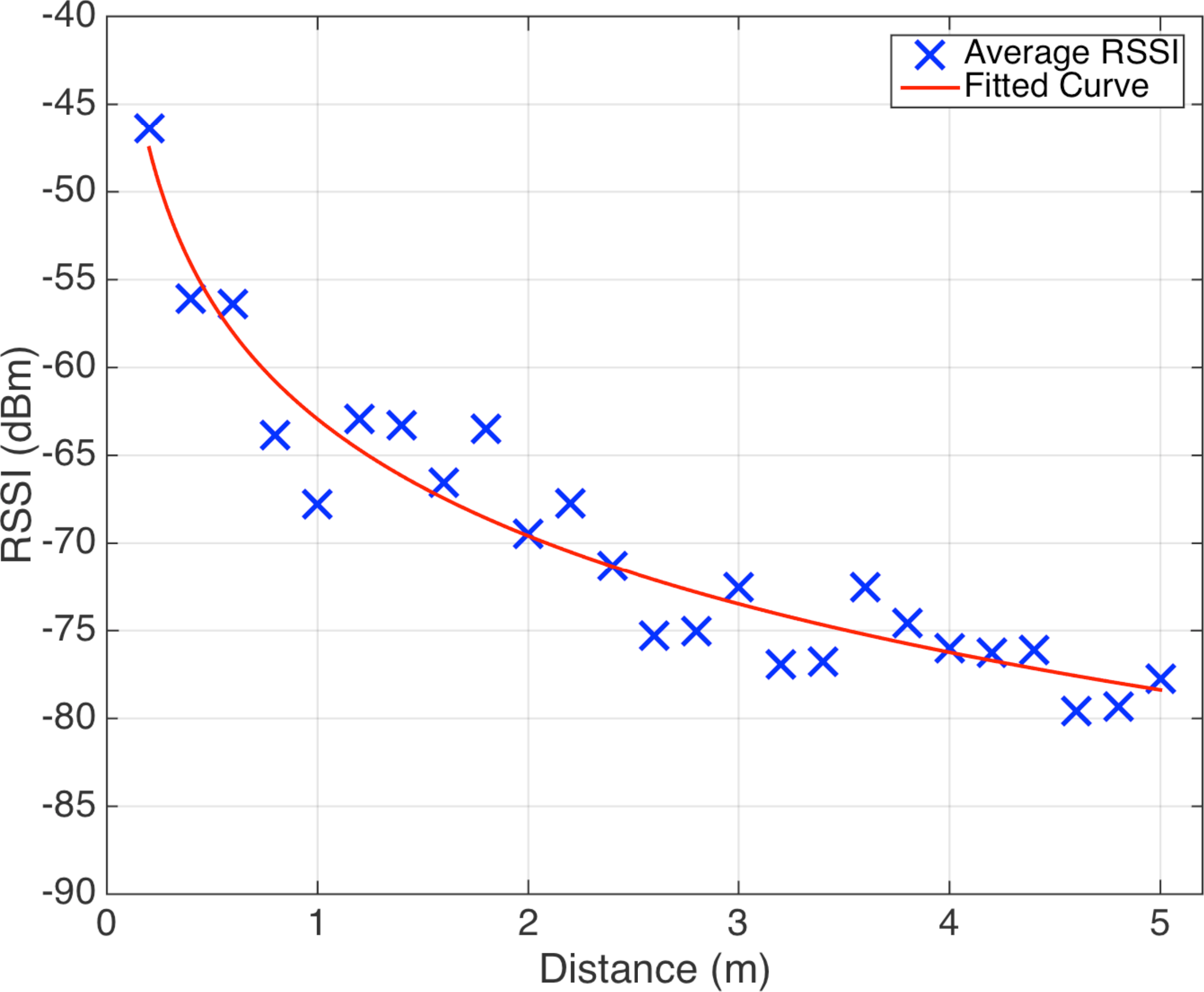}
\label{curve1}}\vspace{1ex}

\subfloat[Corridor.]{\includegraphics[width=0.68\columnwidth]{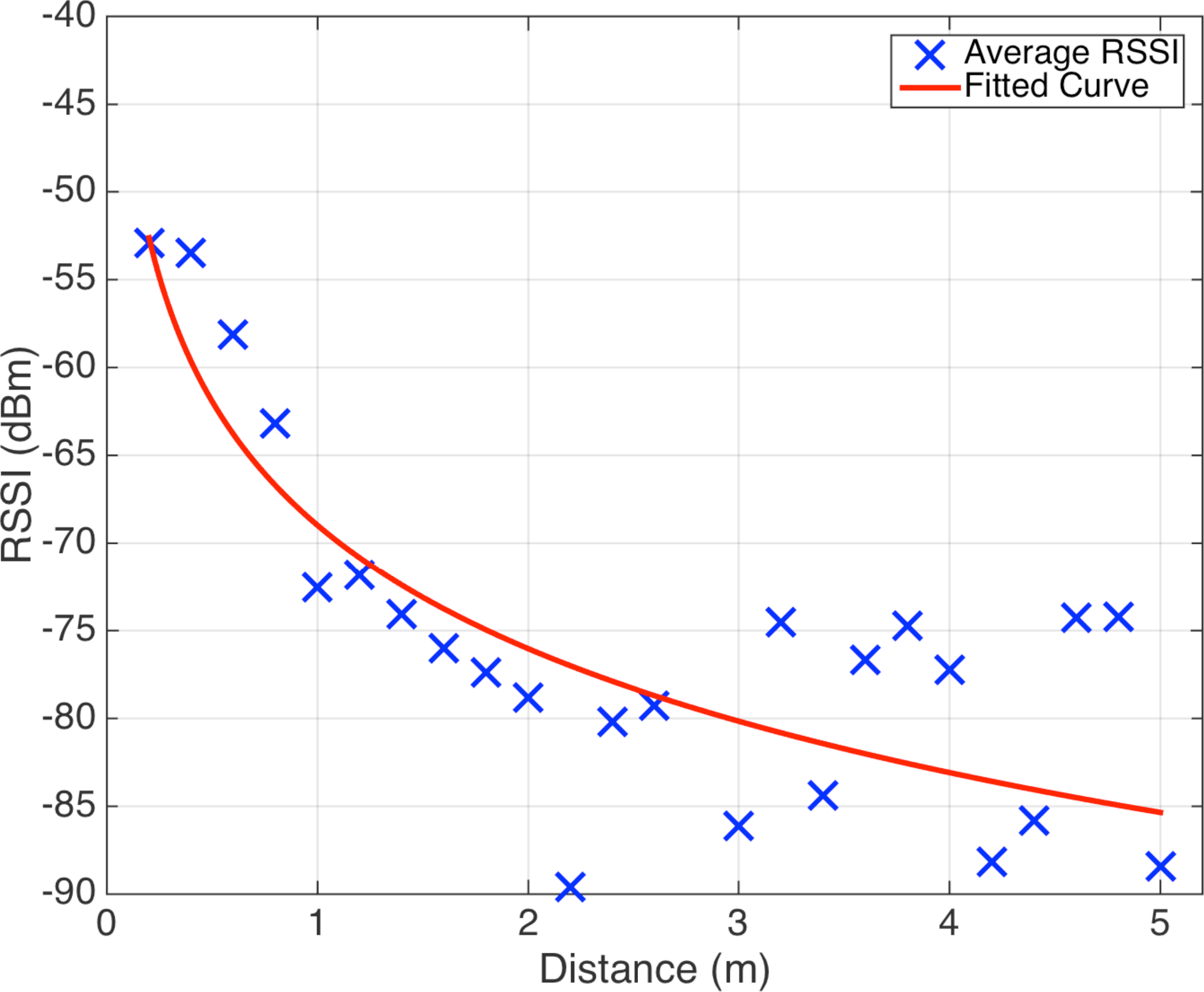}
\label{curve2}}
\caption{Curve fitting for the path loss.}
\label{labraw}
\end{figure}

 \begin{figure}[t!]
\centering 
\captionsetup[subfloat]{farskip=0pt}%
\subfloat[Laboratory.]{\includegraphics[width=0.8\columnwidth]{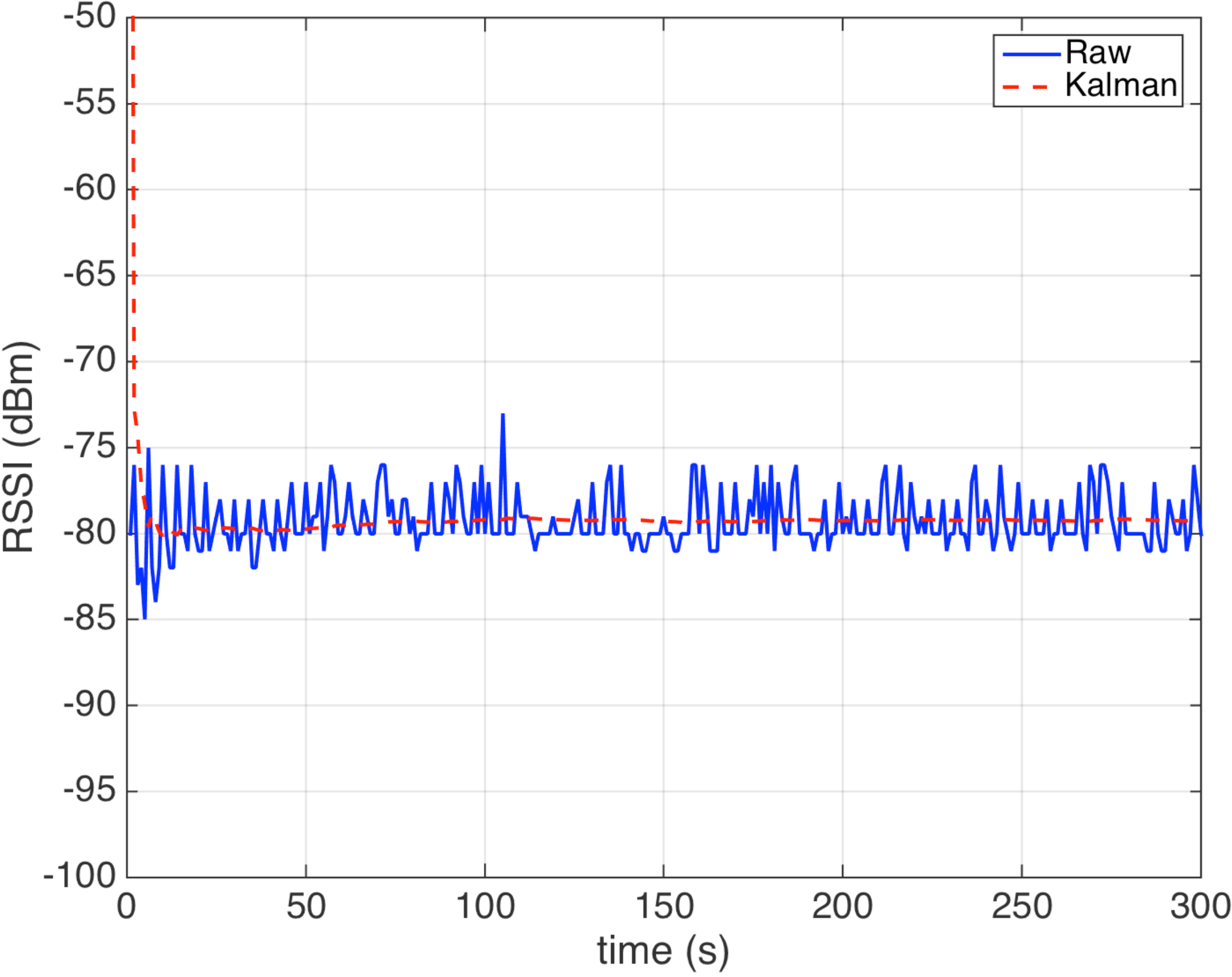}
\label{kalman1}}\vspace{1ex}
\subfloat[Corridor.]{\includegraphics[width=0.8\columnwidth]{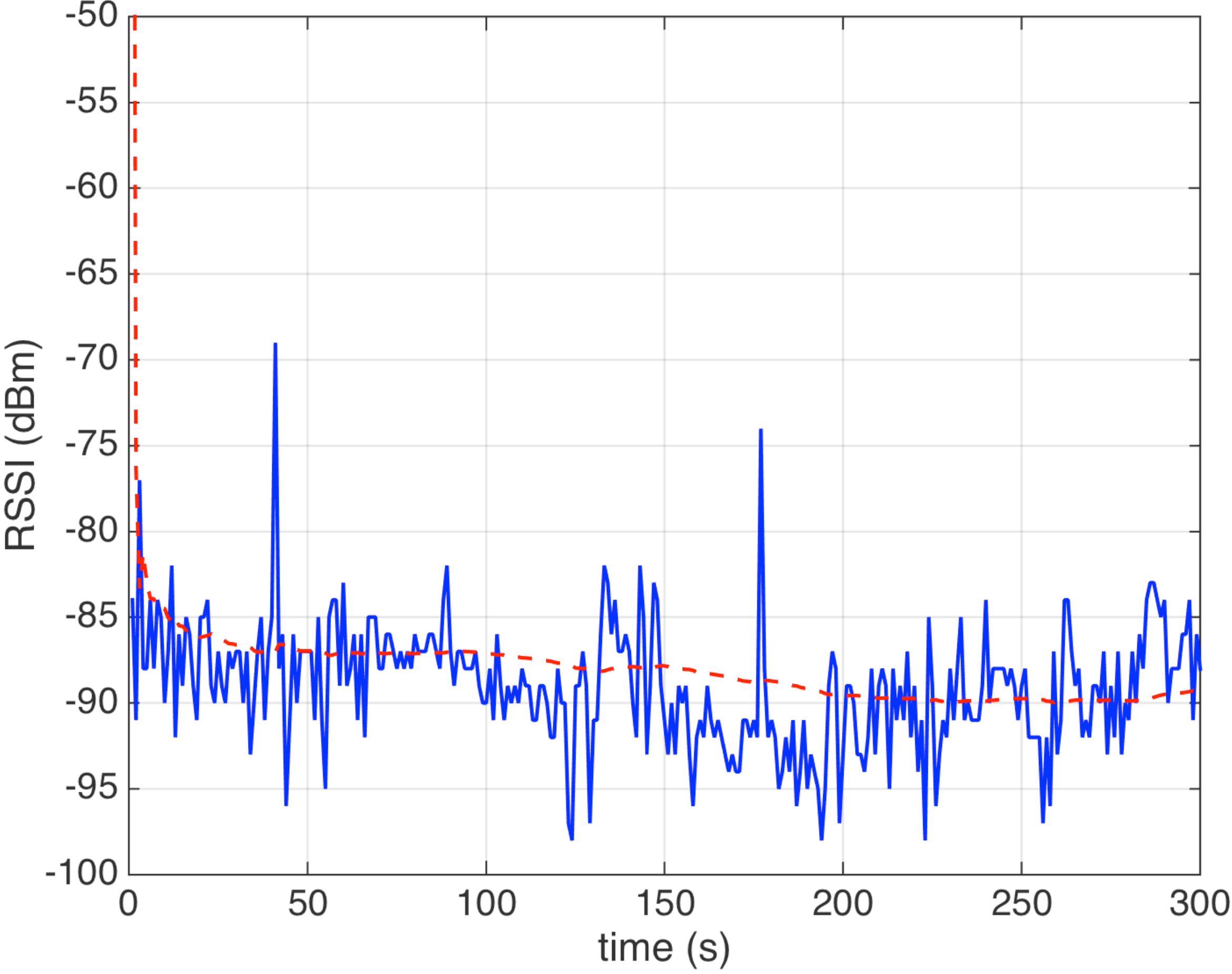}
\label{kalman2}}
\caption{RSSI values and Kalman filter in the two environments in 3~m distance from the beacon.}
\label{kalman}
\end{figure}

When all the RSSI values were collected, a curve fitting for the path loss was performed for the laboratory and the corridor,  shown in Fig.~\ref{curve1} and Fig.~\ref{curve2},  respectively. It is clear that the two environments  experience noise and interference due to other beacon transmissions in the area as well as the general construction of the environment and people's movement. The path loss component, $n$ in Eq.~(\ref{rssimain}), for the laboratory was calculated as $n=2.208$ with $A=-68.99$ and the corridor as $n=2.341$ with $A=-62.94$. According to the experimental results for the path loss, the corridor is a more challenging environment since it seems to be more difficult to follow the fitting curve  and the path loss component is higher. Hence, it is expected that the performance of the system, in terms of accuracy, will be lower in the corridor.

 \begin{figure}[t!]
\centering
\captionsetup[subfloat]{farskip=0pt}%
\subfloat[Laboratory.]{\includegraphics[width=0.75\columnwidth]{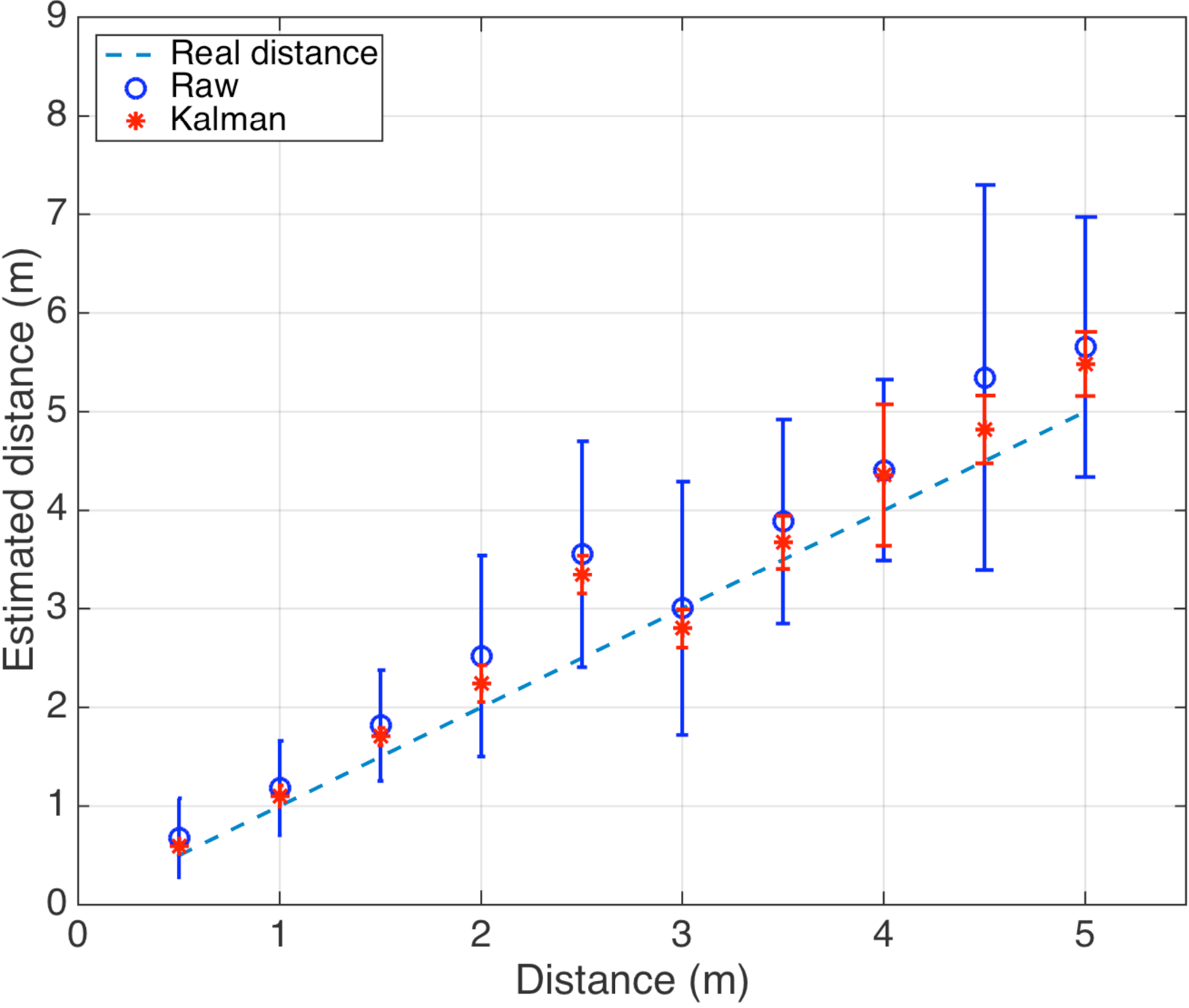}
\label{dis1}}\vspace{1ex}
\subfloat[Corridor.]{\includegraphics[width=0.75\columnwidth]{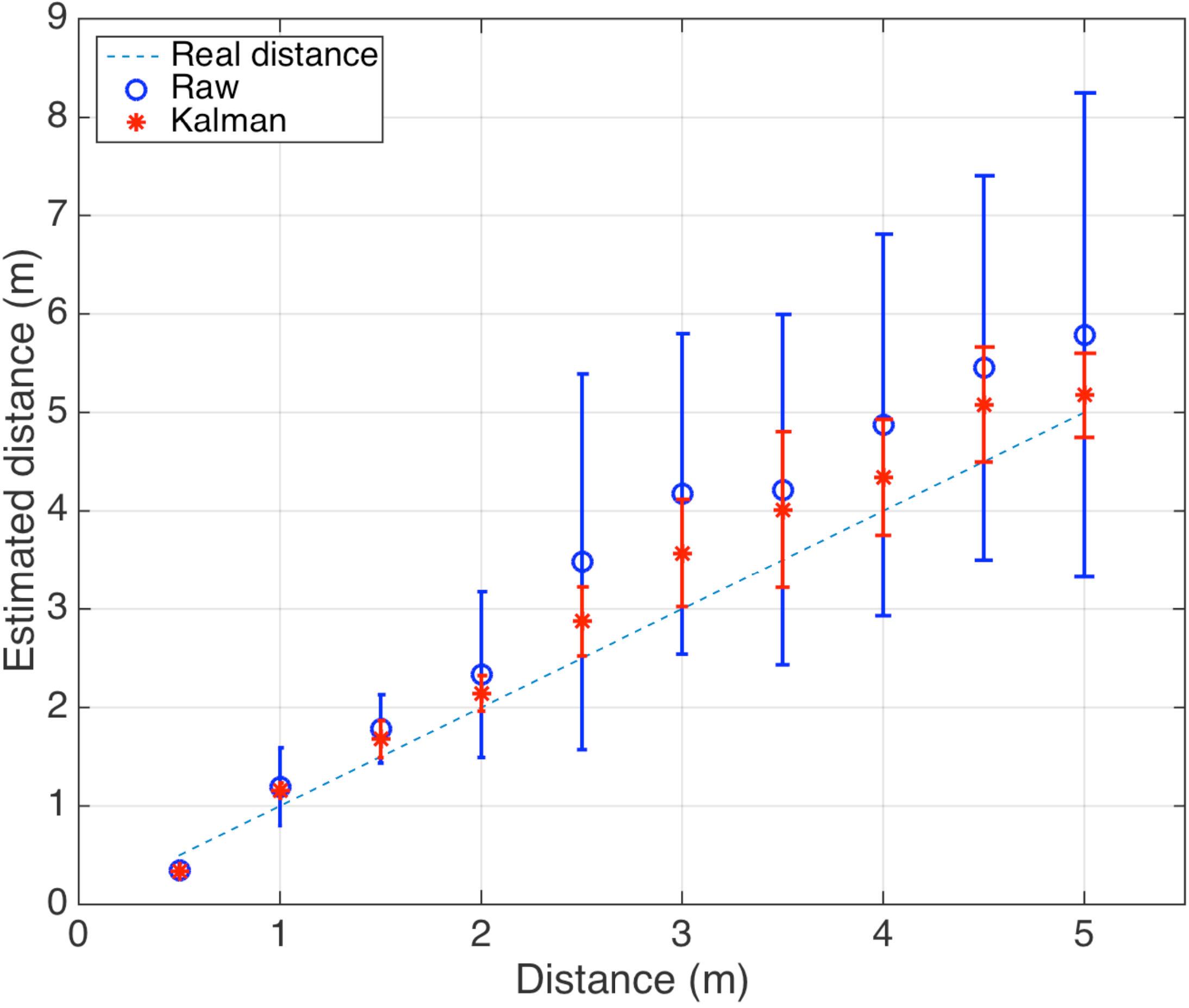}
\label{dis2}}
\caption{Distance estimation performance.}
\label{proximity} 
\end{figure}

\subsection{Proximity performance}
To evaluate the proximity performance,  one Gimbal Series 21 was acting as the transmitter and  a Nexus 5 smartphone was acting as the receiver. The beacon had a transmission interval of 100~ms  and the smartphone was placed in ten distances, in a straight line from the beacon, starting from 50~cm up to 5~m, increasing 50~cm every time. In every distance, more than 100 RSSI measurements were recorded on the smartphone. 

\begin{figure}[t!]
\centering
\captionsetup[subfloat]{farskip=0pt}%
\subfloat[Laboratory.]{\includegraphics[width=0.7\columnwidth]{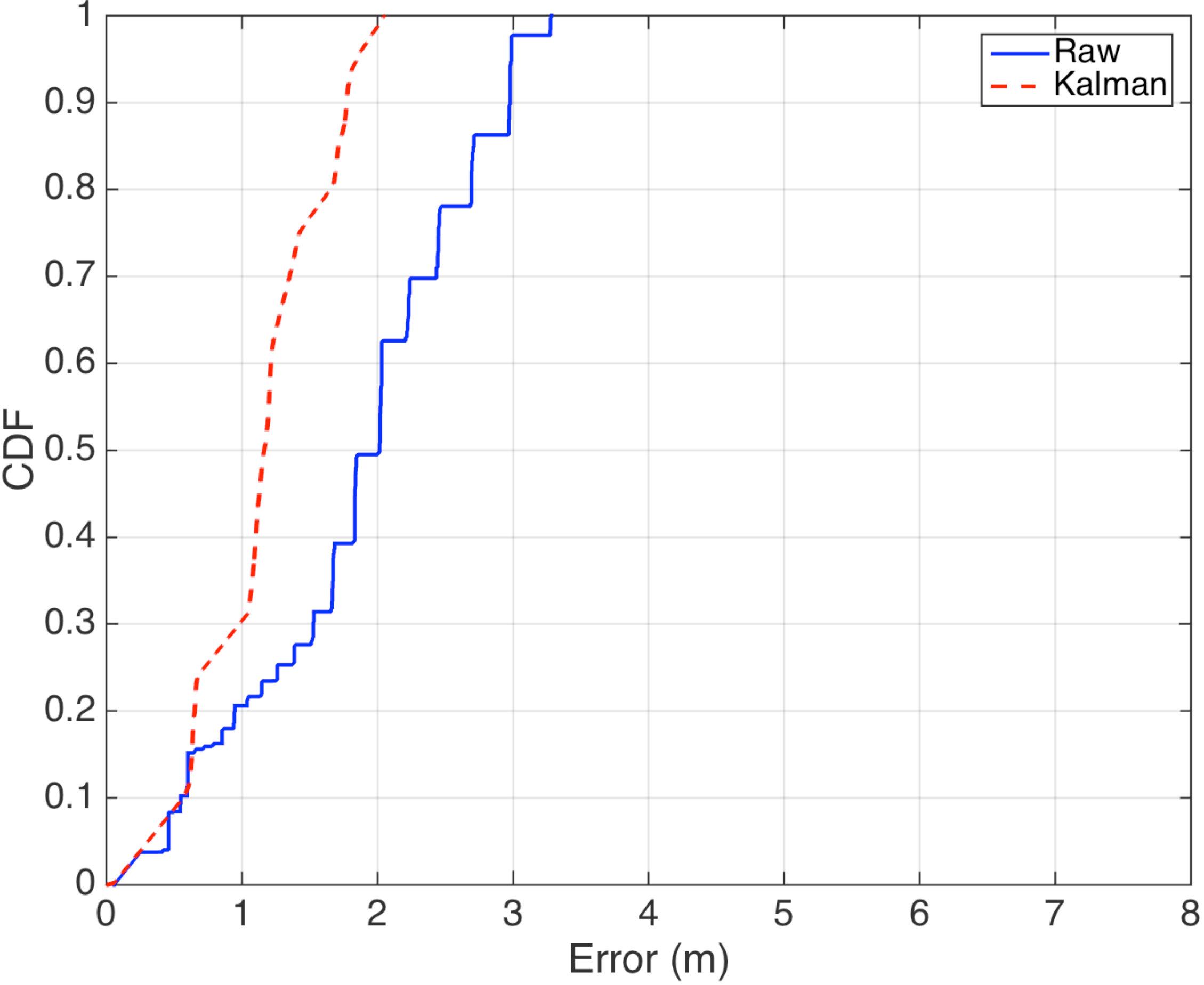}
\label{err1}}\vspace{1ex}
\subfloat[Corridor.]{\includegraphics[width=0.7\columnwidth]{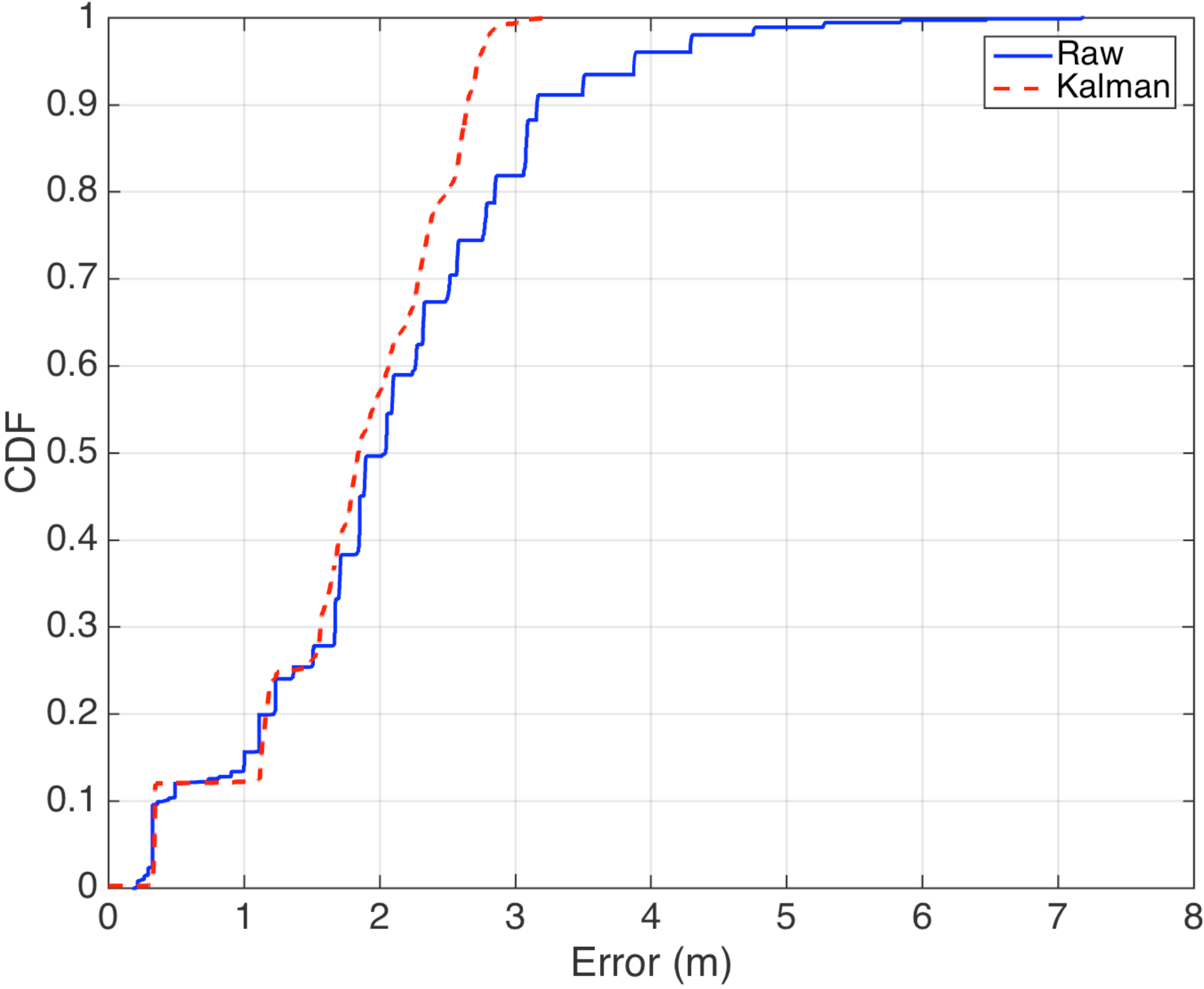}
\label{err2}}
\caption{Cumulative probability error for distance estimation.}
\label{error} 
\end{figure}
               
 \begin{figure}[t!]
\centering
\captionsetup[subfloat]{farskip=0pt}%
\subfloat[Laboratory.]{\includegraphics[width=0.7\columnwidth]{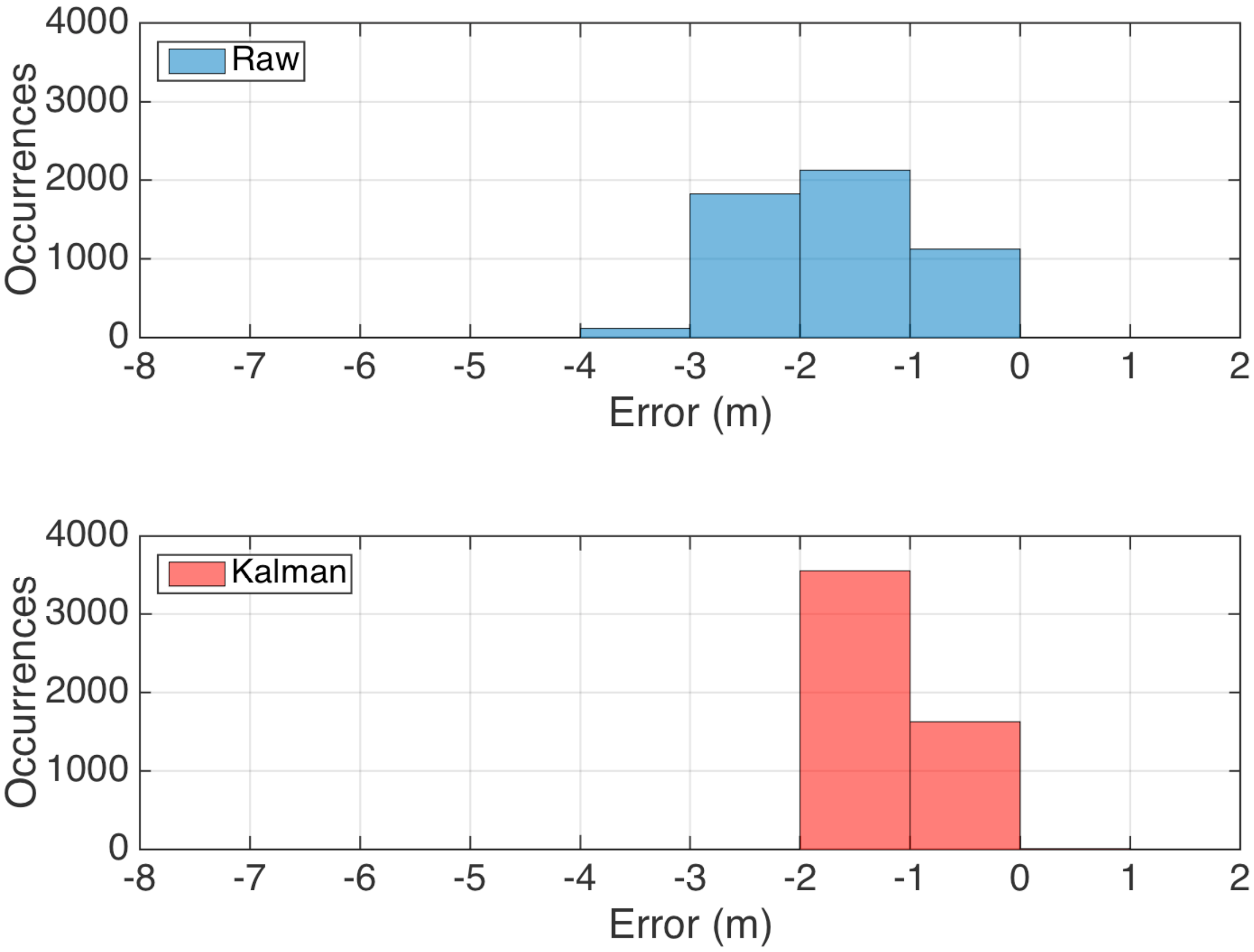}
\label{errb1}}\vspace{1ex}
\subfloat[Corridor.]{\includegraphics[width=0.7\columnwidth]{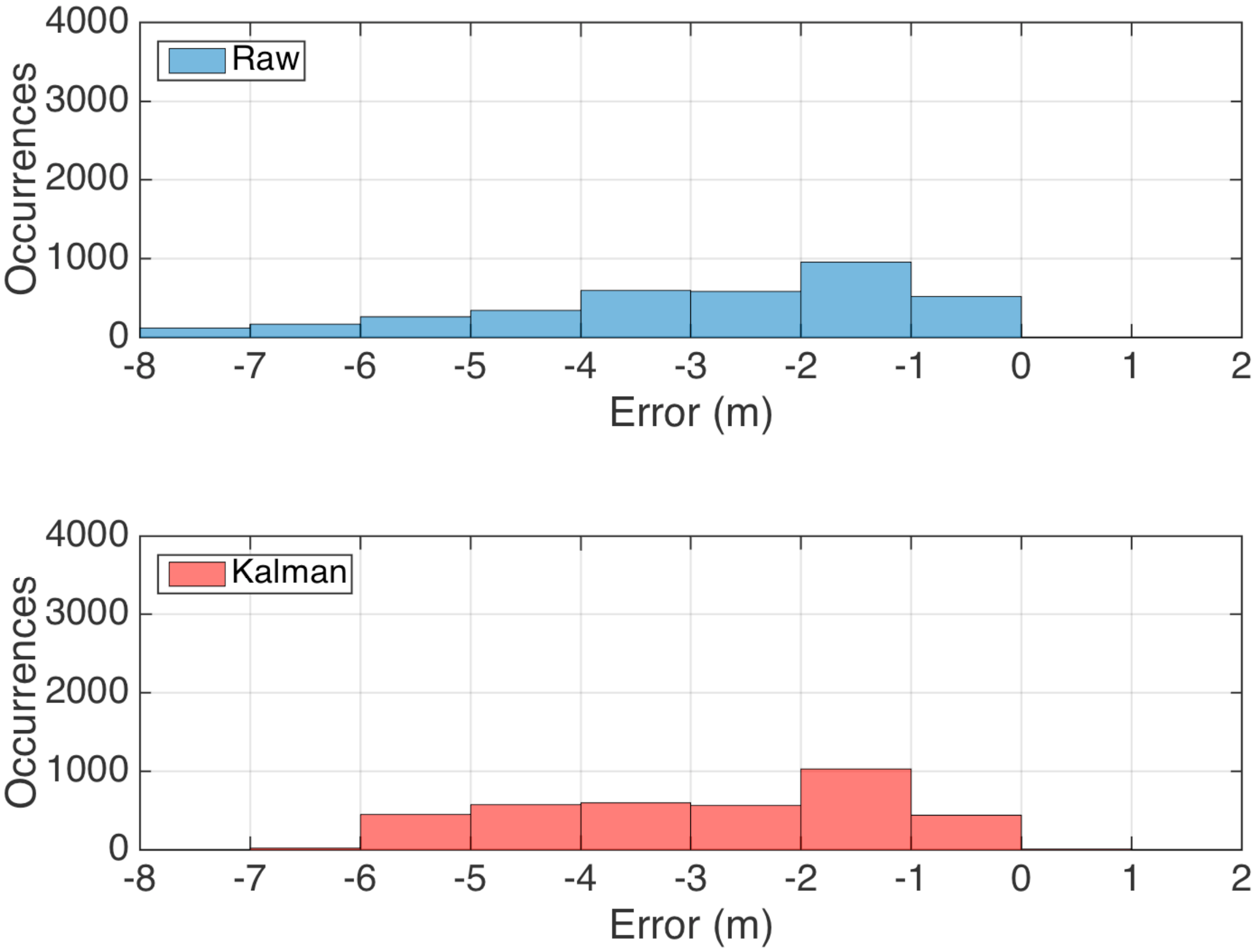}
\label{errb2}}
\caption{Histogram of distance estimation error.}
\label{proximity}
\end{figure}

To improve the accuracy of the system, a Kalman filter was also implemented, similar to~\cite{mackey}. Kalman filter can improve the estimation, especially in noisy environments, such as the two experimental areas. An example of the Kalman filter at different distances in the two environments is shown in Fig.~\ref{kalman}.  It is clear that Kalman filter helps in both environments. In the laboratory, shown in Fig.~\ref{kalman1}, the variation of the RSSI around the mean values is small, so the Kalman filter has a similar value to the final average value. In the corridor, shown in Fig.~\ref{kalman2}, where the environment is more complex in terms of obstacles, material, people moving around and available space, the Kalman filter helps to minimize the effect of the noise.

The distance estimation, using the path loss model for each environment, based on the raw RSSI data and after using Kalman filter for the laboratory and the corridor is shown in Fig.~\ref{dis1} and Fig.~\ref{dis2}, respectively. In general, as the distance of the smartphone from the beacon increases, the RSSI values decrease. This is expected since the signal becomes weaker and it gets affected by other factors such as the material of the objects at the environment and the noise. Another interesting insight from this experiment has to do with the distribution of the values. As the smartphone is moving away from the beacon, the RSSI values have greater variation from the mean value. RSSI is affected  by the environmental parameters, hence, as the distance between the communicating devices increases, the variation of RSSI values increases.

The overall cumulative error  for each environment is shown in Fig.~\ref{err1} and Fig.~\ref{err2}. For the laboratory, the estimation error is less than 3~m, for 95\% of the time, when raw data are used, and for the corridor, it is less than 3.5~m for 95\% of the time. When Kalman filter is used, it can reduce the estimation error. For the laboratory, when Kalman filter is used, the error is within 2~m and for the corridor within 2.5~m. It is clear that Kalman filter can improve the estimation and the overall performance of the proposed system.

A histogram of the distance estimation error for all the distances is shown in Fig.~\ref{proximity}.  For the laboratory, when the error is within 3~m both raw data and Kalman filter have similar performance. When the error is higher than 3~m, Kalman filter helps to improve the system performance by smoothing the values based on previous RSSI values. Similarly, for the corridor, when the error is within 6~m, the performance is the same, while after 6~m, Kalman filter minimizes the errors.

\subsection{Localization performance}
\begin{figure}[t!]
    \centering
    \includegraphics[width=0.8\columnwidth]{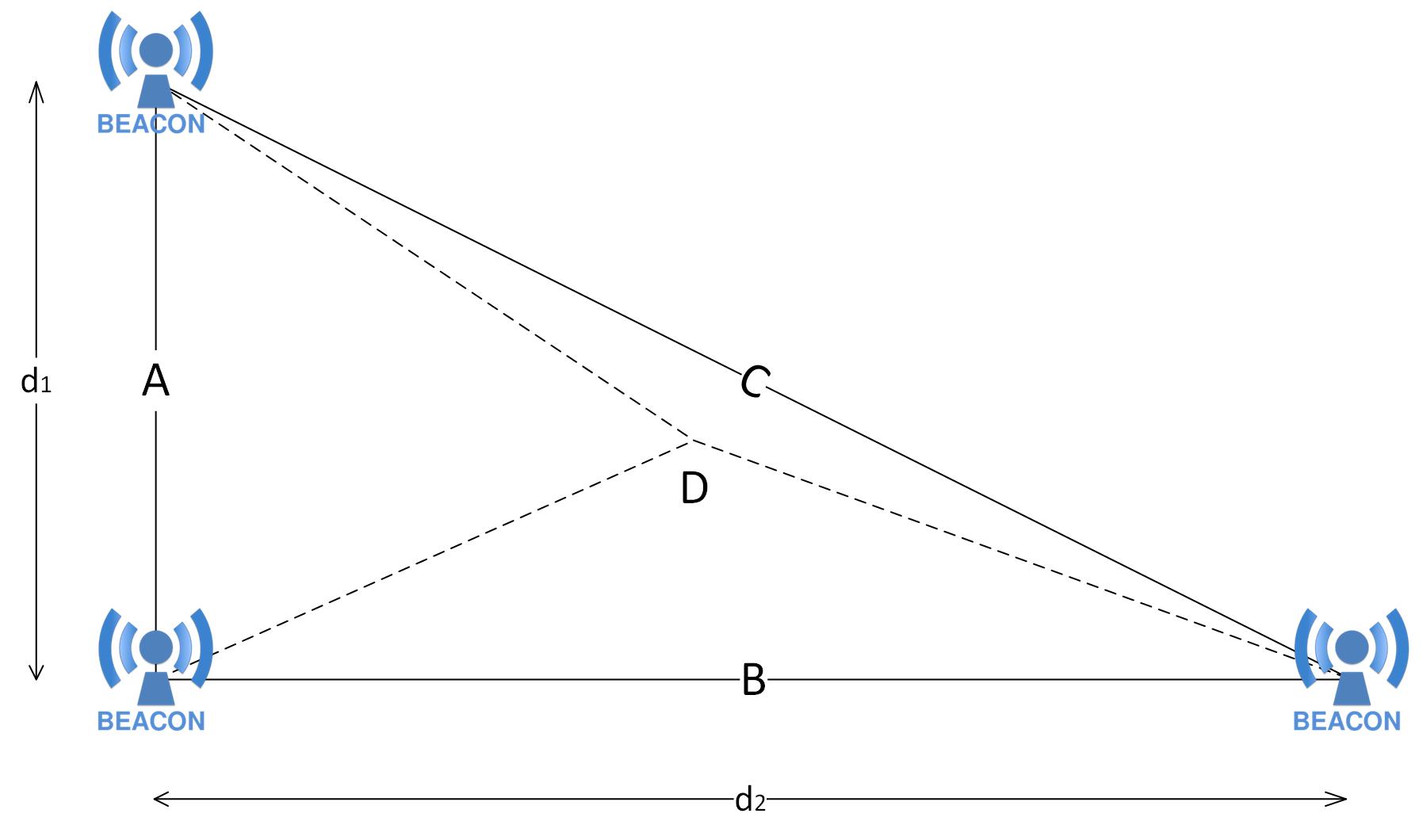}%
     \caption{Localization experiment topology.}
 \label{triexp}
 \end{figure}

The localization performance of the proposed system was examined in another experiment. Since the size of the corridor was limited, the localization experiment was conducted only in the laboratory. The experimental topology is shown in Fig.~\ref{triexp}. 

 \begin{table}[t!] 
\normalsize
    \centering
    \begin{tabular}{|cc|cccc|}\hline
    
        \multirow{ 2}{*}{d$_{1}$} &  \multirow{ 2}{*}{d$_{2}$} &\multicolumn{4}{c|}{Estimation Error (m)} \\
        & & A & B & C & D  \\ \hline  \hline
        1 & 2 & 0.142 & 0.301 & 0.396 & 0.401 \\ \hline
        3 & 4 & 0.703 & 0.798 & 0.814 & 1.122 \\ \hline

      \end{tabular}
   \caption{Localization estimation error.}
            \label{locerr}

\end{table}

%\begin{minipage}{\textwidth}
% 
%  \begin{minipage}[b]{0.49\textwidth}
%    \centering
%    \includegraphics[width=\columnwidth]{trilateration2.jpg}%
%    \captionof{figure}{Localization experiment topology}
%    \label{triexp}
%  \end{minipage}
%  \hfill
%  \begin{minipage}[b]{0.49\textwidth}
%    \centering
%    \begin{tabular}{|cc|cccc|}\hline
%      
%        \multirow{ 2}{*}{d$_{1}$} &  \multirow{ 2}{*}{d$_{2}$} &\multicolumn{4}{c|}{Estimation Error (m)} \\
%        & & A & B & C & D  \\ \hline  \hline
%        1 & 2 & 0.155 & 0.342 & 0.414 & 0.446 \\ \hline
%        3 & 4 & 0.719 & 0.867 & 0.923 & 1.107 \\ \hline
%      \end{tabular}
%      \captionof{table}{Localization estimation error.}
%            \label{locerr}
%
%    \end{minipage}
%    
%  \end{minipage}

Three  Gimbal Series 21 beacons were used to create a triangular and they had LOS between them. The smartphone was placed in four locations, A, B, C and D for approximately one minute. Then, the location estimation was calculated. The experiment was conducted for two identical topologies with different d$_1$ and d$_2$ distances: at the first experiment the distances were d$_1$ = 1~m and d$_2$ = 2~m while at the second the distances were d$_1$ = 3~m and d$_2$ = 4~m. The distances were selected to be similar to distances between the exhibits in a museum. 
 
The localization estimation error for each location is shown in Table~\ref{locerr}.  In the first topology, where the beacons are close to each other, the estimation error is close. It is interesting to mention that the error is within centimeters from the real distance, although no filtering was used. In the second topology, where the distance between the beacon increases, the estimation error is below 1~m.

In both topologies, when the beacons are close to each other, such as in location A, the estimation has the best performance. On the other hand,  when the smartphone is is equal distance from all three beacons, such as in location D, the estimation has the worst performance, probably due to interference.

The experimental results showed that it is easier to estimate the location of the receiver when it is between two beacons. At the same time, as the distance between the beacons increases, the localization error increases as well.  This is probably due to interference that makes the RSSI signals weak and the estimation poor. Another interesting insight is that when the receiver moves among all the three beacons, the localization is a bit more challenging. This can be due to the interference the beacons create to each other when they transmit their signal.

Another promising insight is that the error is below 1~m in almost all the experiments. Considering the simple topology and the LOS between the beacons, this number might increase in a more complex scenario. However, these are the estimation based on the raw data, hence, advanced filtering techniques can be applied to improve the readings from the beacons.

  \begin{figure}[t!]
    \includegraphics[width=\columnwidth]{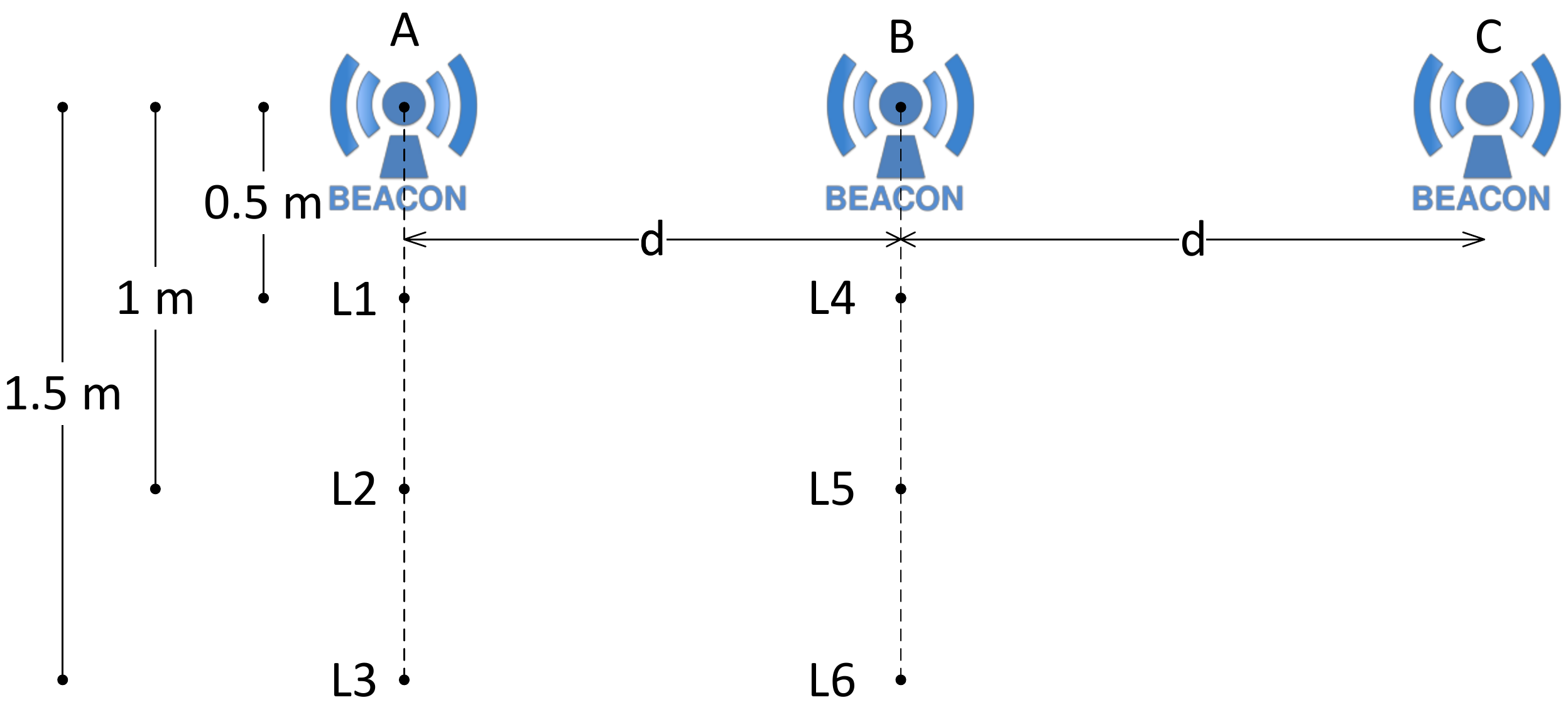}%
    \captionof{figure}{Laboratory topology.}
    \label{estim1}
  \end{figure}
  
  \begin{table} [t!]
    \centering
    \normalsize
      \begin{tabular}[b]{|c|c|c|c|c|c|}\hline
           \multirow{ 2}{*}{d}  &  \multirow{ 2}{*}{Location}& \multicolumn{3}{c|}{Estimation}  &\multirow{ 2}{*}{Accuracy (\%)} \\ 
              &  &  A&   B&  C& \\ \hline

           \multirow{ 6}{*}{1} & L1 &  \cellcolor{Gray}109 & 2  & 0   & \textbf{98.2} \\ 
            & L2 & \cellcolor{Gray}101 &  4 &  1 & \textbf{95.28}  \\ 
            & L3 &  \cellcolor{Gray}100 &  7 & 2  & \textbf{91.74} \\ \cline{2-6}
           
	  & L4 &  6 &  \cellcolor{Gray}101 & 8 & \textbf{87.83}   \\ 
	  & L5 &  9 &   \cellcolor{Gray}109 & 11  &  \textbf{84.5}  \\ 
           & L6 & 10 & \cellcolor{Gray}85  &  16 & \textbf{76.58}  \\  \hline \hline

             \multirow{ 6}{*}{1.5} & L1 &  \cellcolor{Gray}104 & 1  & 0   & \textbf{99.05} \\ 
            & L2 & \cellcolor{Gray}105 &  3 &  1 &  \textbf{96.33}\\ 
            & L3 &  \cellcolor{Gray}102 &  1 & 2  & \textbf{92.73} \\  \cline{2-6}
            
             & L4 &  4 &  \cellcolor{Gray}103 & 8   & \textbf{89.57}  \\ 
             & L5 &  5 &   \cellcolor{Gray}107 & 10  & \textbf{87.7} \\ 
             & L6 & 9 & \cellcolor{Gray}85  &  14 & \textbf{78.7} \\  \hline \hline
            
             \multirow{ 6}{*}{2} & L1 & \cellcolor{Gray} 103  & 0  & 0   & \textbf{100}  \\ 
            & L2 & \cellcolor{Gray}103 &  2 &  0 & \textbf{98.10}  \\ 
            & L3 & \cellcolor{Gray} 102 &  4 & 2  & \textbf{94.44} \\  \cline{2-6}

 	 & L4 &  3 &  \cellcolor{Gray}111 & 6  & \textbf{92.5}  \\
	  & L5 &  4 &  \cellcolor{Gray} 103 & 6  &  \textbf{91.15} \\ 
           & L6 & 9 & \cellcolor{Gray}87  &  14 &  \textbf{79.09} \\  \hline 
            
            \end{tabular}

      \captionof{table}{Laboratory estimation accuracy results.}
\label{tab1}
    \end{table}

\subsection{Detection accuracy}
\begin{figure}
  \centering
    \includegraphics[width=0.6\columnwidth]{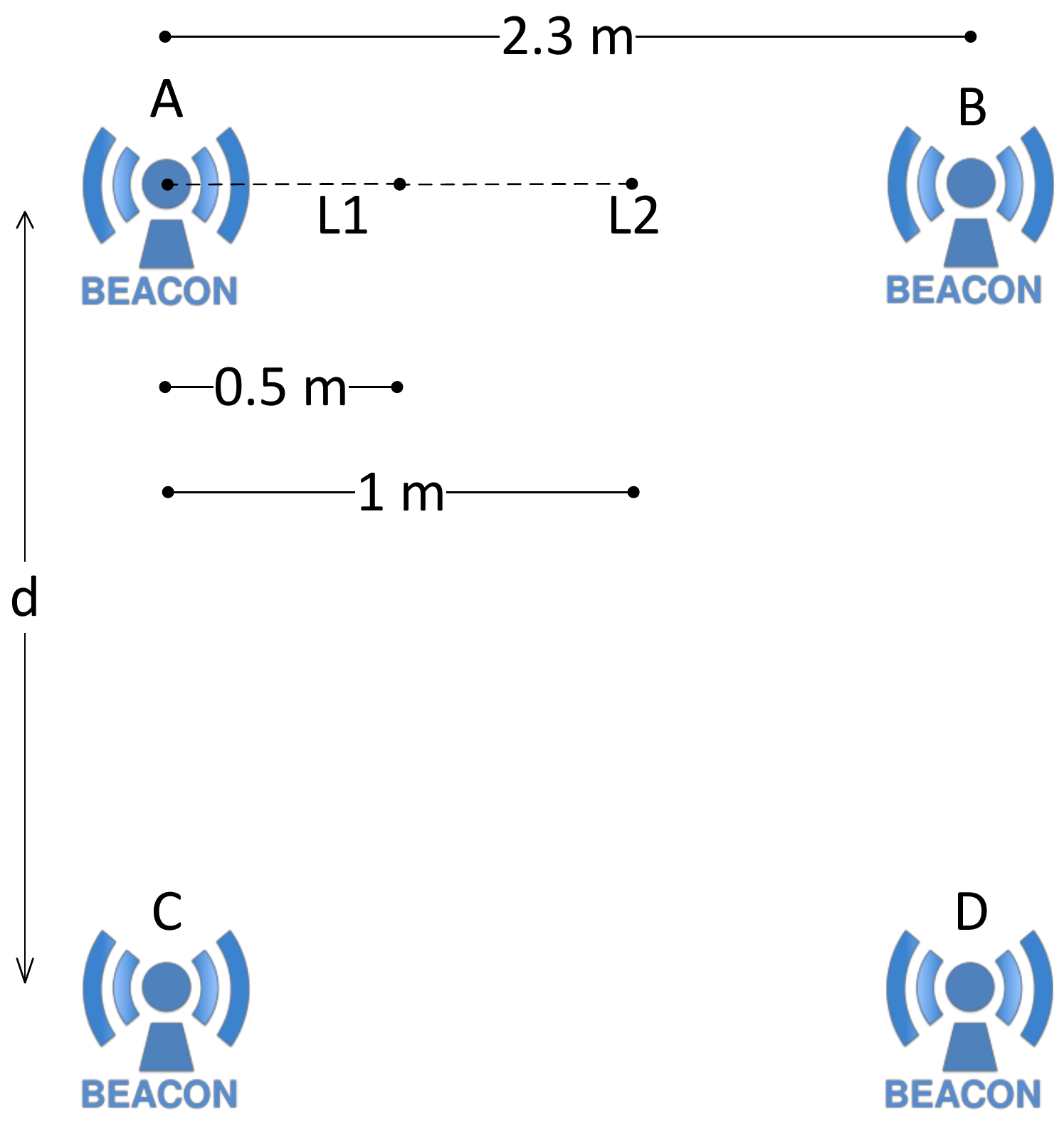}%
    \captionof{figure}{Corridor topology.}
    \label{estim2}
  \end{figure}

The third experiment examines the  detection accuracy of the proposed systems when multiple beacons are in close proximity in the area. Three Gimbal Series 21 beacons  were used with different distances, $d$,  between them.  For the laboratory, topology is shown in Fig.~\ref{estim1} was used, and for the corridor, topology shown in Fig.~\ref{estim2} was used with the detailed results in Table~\ref{tab1} and Table~\ref{tab2}, respectively. Since the second topology was at the corridor, one of the distances  between the beacons was 2.3~m, which is the width of the corridor. 
  
In the laboratory experiment, the smartphone was placed in six different locations, L1- L6, which are in different distances, in vertical connection with beacon A, for locations L1- L3, and  with beacon B, for locations L4- L6, as shown in Fig.~\ref{estim1}. In every location, the application on the smartphone collected  approximately 100 readings and the application predicts the closer beacon, based on the smallest RSSI value. The results along with the estimation accuracy are shown in  Table~\ref{tab1}.

When the distance between the beacons is 1~m, and the receiver is on one side of the topology, close to A, the performance is high. The closer is the receiver to the beacon, the higher the accuracy. As the receiver is moving further from the beacon the accuracy drops. This is due to the reception of signals from the other two beacons with similar RSSI values as of the RSSI values from beacon A. As the distance between the beacons increases, the system accuracy increases as well. When the distance is greater between the beacon and the neighboring devices which create interference, it is easier for the receiver to estimate the beacon that it is closer to.

  \begin{table} 
    \centering
    \normalsize
      \begin{tabular}[b]{|c|c|c|c|c|c|c|}\hline
           \multirow{ 2}{*}{d}  &  \multirow{ 2}{*}{Location}& \multicolumn{4}{c|}{Estimation}  &\multirow{ 2}{*}{Accuracy (\%)} \\ 
              &  &  A&   B&  C& D & \\ \hline

           \multirow{ 2}{*}{1} & L1 &  \cellcolor{Gray}92 & 13  & 8 &3   & \textbf{79.31} \\ 
            & L2 & \cellcolor{Gray}86 &  14 &  9 &2 & \textbf{77.48}  \\   \hline \hline

             \multirow{ 2}{*}{1.5} & L1 &  \cellcolor{Gray}101 & 13  & 4   & 1 & \textbf{84.87} \\ 
            & L2 & \cellcolor{Gray}89 &  14 &  7 &1 &  \textbf{80.18}
        \\  \hline \hline
            
             \multirow{ 2}{*}{2} & L1 & \cellcolor{Gray} 98  & 12  & 2 &0   & \textbf{87.5}  \\ 
            & L2 & \cellcolor{Gray}93 &  12 &  3 &0 & \textbf{86.11}  
            \\  \hline 
            
            \end{tabular}
      \captionof{table}{Corridor estimation accuracy results.}
           \label{tab2}
    \end{table}

When the same experiment is repeated and the beacon is in the middle of the topology, closer to B but in the same distance from A and C, the accuracy drops. As the beacon is moving further from B, the detection of signals from A and C increases and the accuracy of the system drops. As the distance between the beacons increases, that helps the receiver to improve the detection estimation.

It is clear that when the receiver is close to a beacon, within 50~cm, the accuracy of the system is high. As the receiver is moving further away from the beacon, the accuracy drops. At the same time, the distance from other beacons in the area also affects the performance. When the neighboring beacons are within 1~m, the accuracy of the system drops. The accuracy can increase if the distance between the beacons increases. In a museum scenario, when there is sufficient distance between the beacons, and there is minimum interference among them, the application estimation would have an acceptable performance.

At the corridor topology, shown in Fig.~\ref{estim2}, the detection accuracy is lower, as shown in Table~\ref{tab2}. This is expected since the higher noise in this environment. Again, the closer to the beacon the higher the detection accuracy at the receiver. However, as the number of the neighboring devices increases that affect the detection performance. The more beacons in the area, the worst the detection estimation. If the distance between the neighboring beacons increases, that can help to improve the detection estimation, as shown in the experimental results. System performance increases, as the distance between the beacons increases.

\subsection{Discussion}
According to experimental results, beacons have a promising performance for both proximity and localization services. However, their performance based on raw data, without any filtering, might not be sufficient for applications that need beacons to be placed too close to each other. Advanced filtering techniques can be used to improve their performance. If there is Wi-Fi connection in the museum, the filtering can take place in real-time and improve even more the performance of the services.

Another interesting insight from the experiment is that the beacons can drop their proximity and localization accuracy rapidly, when there is a dynamic change in the environment, such as noise and interference. In every application, the deployment area should be studied in advance to improve the estimation of the beacons and minimize the errors.

The main advantage of the proposed system is its simplicity, low cost, and ease of installation. All the experiments took place without creating any interference to the other wireless infrastructures in the area, while the developed Android application required minimum interaction with the user. After installing the application, the users were moving around and when they approached a beacon they get a notification. At the same time, the beacons can be moved around and relocated easily to follow the needs of the exhibits and the collections. The architecture presented in this paper is scalable and can be applied to museums with different sizes.

\section{Conclusions}\label{concl}
In this paper, a BLE localization technology for an IoT-Based smart museum was presented. The proposed system uses BLE beacons to improve the interaction in a museum. iBeacons were used to provide proximity and localization services. An Android application was also developed to examine the performance of the system. The system was designed to work without Internet connection, however, its accuracy can be increased if access to a cloud server is provided.

Three experiments were conducted to measure the performance under different scenarios. The experimental results are promising. BLE beacons can improve the interaction in a museum with low cost and without creating any interference with other wireless infrastructures in the area.

\bibliographystyle{IEEEtran}

\bibliography{IEEEabrv,reference}

\begin{IEEEbiography}
[{\includegraphics[width=1in,height=1.25in,clip,keepaspectratio]{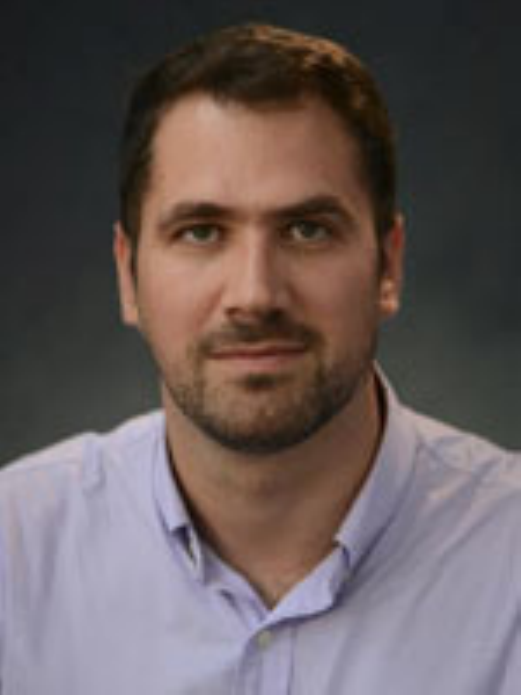}}]{Petros Spachos} (M'14--SM'18) received the Diploma  degree in Electronic and Computer Engineering from the Technical University of Crete, Greece, in 2008, and the M.A.Sc. degree in 2010 and the Ph.D. degree in 2014, both in Electrical and Computer Engineering from the University of Toronto, Canada.  He was a post-doctoral researcher at University of Toronto from September 2014 to July 2015. He is currently an Assistant Professor in the School of Engineering, University of Guelph, Canada. His research interests include experimental wireless networking and mobile computing with a focus on wireless sensor networks, smart cities, and the Internet of Things. He is a Senior Member of the IEEE.
\end{IEEEbiography}

\begin{IEEEbiography}[{\includegraphics[width=1in,height=1.25in,clip,keepaspectratio]{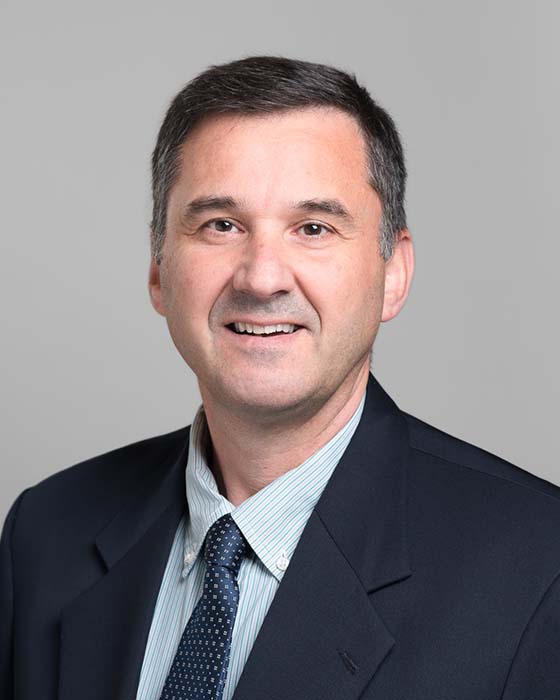}}]{Konstantinos N. (Kostas) Plataniotis}  (S'90--M'92--SM'03--F'12) received his B. Eng. degree in Computer Engineering from University of Patras, Greece and his M.S. and Ph.D. degrees in Electrical Engineering from Florida Institute of Technology Melbourne, Florida. Dr. Plataniotis is currently a Professor with The Edward S. Rogers Sr. Department of Electrical and Computer Engineering at the University of Toronto in Toronto, Ontario, Canada, where he directs the Multimedia Laboratory. He holds the Bell Canada Endowed Chair in Multimedia since 2014. His research interests are primarily in the areas of image/signal processing, machine learning and adaptive learning systems, visual data analysis, multimedia and knowledge media, and affective computing. Dr. Plataniotis is a Fellow of IEEE, Fellow of the Engineering Institute of Canada, and registered professional engineer in Ontario.

Dr. Plataniotis has served as the Editor-in-Chief of the IEEE Signal Processing Letters. He was the Technical Co-Chair of the IEEE 2013 International Conference in Acoustics, Speech and Signal Processing, and he served as the inaugural IEEE Signal Processing Society Vice President for Membership (2014 -2016) and General Co-Chair for the 2017 IEEE GLOBALSIP. He serves as the 2018 IEEE International Conference on Image Processing (ICIP 2018) and the 2021 IEEE International Conference on Acoustics, Speech and Signal Processing (ICASSP 2021) General Co-Chair.
\end{IEEEbiography}

%%%%%%%%%%%%%%%%%%%%%%%%%%%%%%%%%%%%%

%%%%%%%%%%%%%%%%%%%%%%%%%%%%%%%%%%%%%%%%%%
\end{document}